\documentclass[12pt]{article}
\usepackage{amsmath}
\usepackage{graphicx,psfrag,epsf}
\usepackage{enumerate}
\usepackage{url} 
\usepackage{blindtext}
\usepackage{comment}
\usepackage{booktabs}
\usepackage{listings}
\usepackage{amsthm}
\usepackage{xcolor}
\usepackage{siunitx}
\usepackage{algorithm}
\usepackage{algpseudocode}
\usepackage{amssymb}
\usepackage{comment}
\algrenewcommand\algorithmicrequire{\textbf{Input:}}
\algrenewcommand\algorithmicensure{\textbf{Output:}}
\newtheorem{theorem}{Theorem}
\newtheorem{remark}{Remark}
\newtheorem{lemma}{Lemma}
\newtheorem{assumption}{Assumption}
\newtheorem{corollary}{Corollary}
\newtheorem{proposition}{Proposition}
\usepackage{natbib}
\usepackage{subcaption}
\newcommand{\blind}{1}

\date{}
\begin{document}

\def\spacingset#1{\renewcommand{\baselinestretch}%
{#1}\small\normalsize} \spacingset{1}


\if1\blind
{
  \title{\bf Bayesian Mirror for False Discovery Rate Control}
  \author{
  Theodosios Papazoglou$^{*}$
  and 
  Guosheng Yin$^{*}$\hspace{.2cm}\\
  \emph{$^{*}$Department of Statistics and Actuarial Science, University of Hong Kong}
}
  \maketitle
} \fi

\if0\blind
{
  \bigskip
  \bigskip
  \bigskip
  \begin{center}
    {\LARGE\bf Bayesian Mirrors for False Discovery Rate Control}
\end{center}
  \medskip
} \fi

\bigskip
\begin{abstract}
Controlling the false discovery rate (FDR) is a necessary requirement of variable selection in regression. Recently developed mirror methods achieve this by perturbing each predictor in opposite directions and contrasting the two resulting coefficients, giving a mirror statistic that is symmetric about zero under the null. We propose the Bayesian mirror, which characterises the admissible perturbations of each predictor geometrically, as a sphere of uniquely determined radius in the residual space of the remaining predictors. A uniform prior on the sphere, combined with a cut posterior that keeps the perturbations independent of the response, makes the mirror statistic exactly symmetric under the null at every Monte Carlo iteration. Since these spheres share a common subspace across predictors, each draw reduces to a closed-form two-dimensional update requiring a single factorisation of the design.\ The Monte Carlo selections are aggregated through the e-Benjamini--Hochberg procedure, yielding asymptotic global FDR control under a spectral condition on the design that replaces the weak-dependence assumptions of existing mirror methods. We further establish power guarantees, including recovery of the full active set by the aggregated procedure, and extend the method to high dimensions through data-split screening. In simulations and a semi-synthetic study on gene expression data, the Bayesian mirror controls the FDR under strong dependence, where most competing methods fail, with competitive power.

\end{abstract}

\noindent%

\spacingset{1.45} 
\section{Introduction}

Identifying a small number of relevant variables from a large pool of candidates is one of the most ubiquitous inferential challenges in modern data analysis \citep{George2000TheVar}. Classical approaches to model selection (e.g., stepwise regression), penalised methods (e.g., Lasso), and likelihood-based criteria (e.g., AIC and BIC), optimise predictive accuracy or model parsimony, and provide no formal guarantee on the proportion of spurious inclusions in the selected set. The false discovery rate (FDR) of \citet{Benjamini1995}, the expected proportion of null variables among the selected ones, offers a principled and scientifically interpretable error criterion for this task. Controlling it in regression requires a reference distribution for each candidate variable under the null, conditional on the remaining predictors \citep{Barber2015CONTROLLING-THE}.

The procedure of \citet{Benjamini1995} controls the FDR for marginal $p$-values that are independent or, as shown by \citet{Benjamini2001}, positively regression dependent. However, these conditions are not guaranteed in variable selection for regression, since the hypotheses concern conditional effects, so the test statistics inherit the dependence of the design, and any selection or penalisation step alters the null distribution of the estimates. This has motivated methods that embed FDR control within the selection procedure itself, replacing the marginal $p$-values by a statistic whose null behaviour is known by symmetry. The knockoff filter of \citet{Barber2015CONTROLLING-THE} constructs for each predictor a synthetic copy that mimics the correlation structure of the design while being conditionally independent of the response, contrasts each variable with its copy through a statistic whose sign is symmetric under the null, and selects through a data-driven threshold, yielding finite-sample FDR control. 

The framework has since been extended to arbitrary response distributions and dimensions through model-X knockoffs \citep{Candes2018Panning-for-Gol}, to high-dimensional and robust inference \citep{Barber2019A-knockoff-filt,Barber2020RobustInf}, and to knockoffs generated by deep generative models \citep{Romano2020Deep-Knockoffs} and Metropolis-based samplers \citep{Bates2021Metropolized-Kn}. A limitation shared by all single-run implementations is that the knockoffs are generated once at random, so two applications of the same procedure to the same data may select markedly different sets. \citet{Ren2023Derandomizing-K} addressed this by aggregating selection frequencies across multiple runs, without a finite-sample guarantee on the global FDR, and \citet{Ren2024Derandomised-kn} supplied one by aggregating the runs through e-values \citep{Vonk2021E-values} and the e-BH procedure \citep{Wang2022False-Discovery}.

Under strong dependence among the covariates, knockoffs become harder to distinguish from their originals and the power of knockoff-based procedures deteriorates \citep{Xing2023Controlling-Fal}. Mirror statistics retain the sign-flip symmetry while dispensing with a synthetic copy of the design. In the Gaussian mirror of \citet{Xing2023Controlling-Fal}, each predictor is perturbed in opposite directions by a random Gaussian perturbation, and the contrast between the two resulting coefficients is informative only when the predictor is relevant. The symmetry of the mirror statistic under the null requires the two estimated coefficients to be uncorrelated, a condition that \citet{Xing2023Controlling-Fal} enforced algebraically by rescaling the realised draw by a constant computed from the design, which requires a separate fit for each of the $p$ predictors. \citet{Dai2023False-Discovery} instead obtained the symmetry of the mirror statistic by data splitting, at the cost of estimating each coefficient from half of the observations. In both methods the FDR control is asymptotic and rests on a weak dependence condition among the mirror statistics, which both establish for high-dimensional linear models under bounded eigenvalues of the covariance matrix of the predictors. The nullstrap method of \citet{wang2025nullstrapsimplehighpowerfast} discards symmetry and calibrates the selection threshold on a synthetic null response generated from the estimated noise level, so the procedure relies on the coefficients and noise level of a single full-data fit.

These ideas have also been explored within Bayesian paradigms. \citet{Yap2022BayesianVar} showed that knockoff variables can be embedded within Bayesian variable selection, a compatibility briefly noted by \citet{Candes2018Panning-for-Gol}.  
\citet{molinari2025falsediscoveryratecontrol} formed the mirror statistics from posterior draws under mean-field variational inference, avoiding data splitting and covariate perturbation. Their target is the Bayesian FDR, and the null symmetry is a property of the approximating family rather than of the design, with the error control resting on simulation evidence. 
\citet{Gu2021Bayesian-Knocko} proposed a fully Bayesian knockoff filter in which the knockoff variables are regenerated from their prior at each iteration of a Gibbs sampler, preserving conditional independence with the response, and controlling the Bayesian FDR.
 
We propose the Bayesian mirror, a variable selection procedure that reinterprets the Gaussian mirror framework of \citet{Xing2023Controlling-Fal} geometrically. The perturbations under which the two mirror coefficients are orthogonal form a sphere of uniquely determined radius in the residual space of the remaining predictors, and we place a uniform prior on this sphere. Under a cut posterior \citep{Plummer2015-Cuts} that keeps the perturbation independent of the response, the mirror statistic formed from posterior draws is exactly symmetric about zero under the null at every Monte Carlo iterate. The iterates are aggregated into a single selection rule via the e-BH procedure of \citet{Wang2022False-Discovery}.

The construction has three main advantages. First, it is fast. The residual spaces of all predictors share a common subspace, so one factorisation of the design serves every variable and each posterior draw is available in closed form, with no per-predictor refitting and no Markov chain. Second, the FDR guarantee does not rest on the weak dependence condition of existing mirror methods. Global FDR control of the aggregated rule holds asymptotically under a spectral condition on the design, which implies weak dependence and is governed by the partial rather than the marginal correlations among the predictors, so that it accommodates strongly correlated settings. Third, we establish power guarantees absent from the mirror literature, including recovery of the full active set by the aggregated procedure. The method extends to high dimensions through screening on a separate subset of the data. In simulations under strong dependence and in a semi-synthetic study on gene expression data, the Bayesian mirror controlled the FDR where most competing methods failed, with competitive power and the shortest run time among the methods compared.


\section{Bayesian Mirror}
\subsection{Problem Setup}
We observe a response $y\in\mathbb{R}^n$ and a design matrix $X\in\mathbb{R}^{n\times p}$ of full column rank whenever $n>p$, with $j$-th column  $X_j$ and $X_{-j}\in\mathbb{R}^{n\times(p-1)}$ the matrix with column $j$ removed. The data follow a linear regression model, $y=X\beta^0+\epsilon$, $\epsilon\sim N(0,\sigma_0^2I_n)$, with active set $\mathcal{H}_1=\{j:\beta^0_j\neq0\}$, null set $\mathcal{H}_0=\{j:\beta^0_j=0\}$, $p_1=|\mathcal{H}_1|$ and $p_0=|\mathcal{H}_0|$.

We write $\Omega=(X^\top X)^{-1}$, $H=I_n-X(X^\top X)^{-1}X^\top$, and $H_{-j}=I_n-X_{-j}(X_{-j}^\top X_{-j})^{-1}X_{-j}^\top$ for the projection onto the orthogonal complement of $\mathrm{col}(X_{-j})$, whose dimension is $r=n-p+1$. We denote  $\alpha_j=X_j^\top H_{-j}X_j$ the squared norm of the residual of $X_j$ on $X_{-j}$.

For predictor $j$, the mirror variables are $X_j^+$ and $X_j^-$, and the augmented model has coefficient vector $\beta^j=(\beta_j^+,\beta_j^-,\beta_{-j}^\top)^\top$, with nuisance coefficients $\beta_{-j}\in\mathbb{R}^{p-1}$, and noise variance $\sigma_j^2$. Index $0$ refers to the data-generating model and index $j$ to the augmented model for variable $j$.

Throughout, $\|\cdot\|$ is the Euclidean norm, $\|\cdot\|_{\mathrm{op}}$ the operator norm, $|A|$ the cardinality of a set $A$, and $a \vee b=\max(a,b)$. All probability statements are conditional on $X$, with further conditioning indicated by a subscript on $\mathbb{P}$ and $\mathbb{E}$.

\subsection{Frequentist Gaussian Mirror}\label{sec:GM}
For each variable $j=1,\cdots,p$, the Gaussian mirror (GM) method of \cite{Xing2023Controlling-Fal} constructs a pair of mirror variables, $(X_j^+,X_j^-)$, by symmetrically perturbing $X_j$ about a randomly generated direction. Specifically, a perturbation vector $z_j\sim N(0,I_n)$ is drawn independently for each variable, and the mirror variables are defined as
\begin{equation}\label{eq:FGM}
X_j^+ = X_j + c_j z_j, \qquad X_j^- = X_j - c_j z_j,
\end{equation}
where $c_j$ is a scaling constant chosen to ensure the exchangeability of $X_j^+$ and $X_j^-$ under the null. For each $j$ the augmented matrix
\[
\mathbf{X}^j = (X_j^+,\, X_j^-,\,X_{-j}) \in \mathbb{R}^{n \times (p+1)}
\]
is formed and a linear model is fitted, yielding estimated coefficients $\hat{\beta}^j=(\hat{\beta}_j^+,\hat{\beta}_j^-,\hat{\beta}_{-j})$. The mirror statistic for variable $j$ is then defined as
\[
M_j = |\hat{\beta}_j^+ + \hat{\beta}_j^-| - |\hat{\beta}_j^+ - \hat{\beta}_j^-|,
\]
where the sum $\hat{\beta}_j^++ \hat{\beta}_j^-$ carries the effect of $X_j$ and the difference $\hat{\beta}_j^+- \hat{\beta}_j^-$ carries only noise. \citet{Xing2023Controlling-Fal} chose $c_j$ as a function of the realised $z_j$ so that the residualised norm of $c_jz_j$ matches that of $X_j$, which makes $\hat{\beta}_j^+$ and $\hat{\beta}_j^-$ uncorrelated given $X_{-j}$. Under the null $\beta_j^0=0$, $M_j$ is then symmetric about zero, while under the alternative it tends to take large positive values. Variable $j$ is selected if $M_j\ge\tau_q$, where $\tau_q$ is the smallest $s>0$ at which the estimated false discovery proportion does not exceed $q$, as in the knockoff filter of \citet{Barber2015CONTROLLING-THE}. The resulting FDR control is asymptotic, whereas knockoffs control the FDR in finite samples \citep{Barber2015CONTROLLING-THE,Candes2018Panning-for-Gol}. Since each augmented matrix $\mathbf{X}^j$ is specific to variable $j$, the procedure requires $p$ separate model fits, one per variable, and this cost dominates its run time when $p$ is large. 

\subsection{Geometry of Bayesian Mirror Variables}\label{sec:bm_variables}
We first develop the Bayesian mirror for the low-dimensional case where $n>p$. In \citet{Xing2023Controlling-Fal}, the null symmetry rests on an algebraic choice of $c_j$ that makes the two mirror coefficients uncorrelated. Our construction is instead geometric. Each perturbation is drawn from a fixed-radius sphere in the orthogonal complement of the remaining predictors, so that the two mirror variables are exactly orthogonal after residualisation for every admissible perturbation, laying the foundation for the exact null symmetry of the Bayesian mirror statistic. 

First, we define the mirror variables,
\begin{equation*}
    X_j^+=X_j+w_j, \quad X_j^-=X_j-w_j,
\end{equation*}
where the perturbation $w_j$ combines the direction and scaling constant of \eqref{eq:FGM}. To integrate the mirror variables into the likelihood, we specify a prior distribution on $w_j$. Specifically, let $w_j=U_jv_j$ with $U_j\in\mathbb{R}^{n\times r}$ an orthonormal basis of $\mathrm{col}(H_{-j})$ and $v_j\in\mathbb{S}_j=\{v\in\mathbb{R}^{r}:\|v\|=\sqrt{\alpha_j}\}$, i.e.\ the sphere in $r$ dimensions with a radius of $\sqrt{\alpha_j}$. Since $n>p$, this implies $r\ge2$, so $\mathbb{S}_j$ is a continuum and the uniform distribution on it is non-atomic. The design matrix for variable $j$ is then 
\[\mathbf{X}^j=[X_j+w_j,X_j-w_j,X_{-j}]\in\mathbb{R}^{n\times (p+1)}.\]

\begin{lemma}[Orthogonality of mirrors]\label{th:1}
For every $j=1,\dots,p$ and every $w_j\in\mathbb{R}^n$ the mirror variables $X_j^+$ and $X_j^-$ satisfy 
\[
(X_j^+)^\top H_{-j} X_j^- \; = \; 0
\]
if and only if $\|H_{-j}w_j\|^2=\alpha_j$. Equivalently $X_j^+$ and $X_j^-$ are orthogonal after projection onto the orthogonal complement of $\mathrm{col}(X_{-j})$.
\end{lemma}

\begin{remark}
    For $w_j\in\mathrm{col}(H_{-j})$, $H_{-j}w_j=w_j$, so by Lemma~\ref{th:1} the admissible perturbations in the residual subspace are exactly the sphere of radius $\sqrt{\alpha_j}$. The prior support $\mathbb{S}_j$ is therefore the complete set of admissible perturbations within that subspace rather than a convenient choice. The restriction $w_j\in\mathrm{col}(H_{-j})$ also gives $X_{-j}^\top w_j=0$, hence $X_{-j}^\top X_j^+=X_{-j}^\top X_j^-=X_{-j}^\top X_j$, and both mirror variables have the same covariance with the nuisance predictors as $X_j$.
\end{remark}

The radius $\sqrt{\alpha_j}$ coincides with the numerator of the scaling constant $c_j$ of \citet{Xing2023Controlling-Fal}. The construction is determined by the residualised predictors $H_{-j}X_j$, which admit a closed-form in terms of the precision matrix of the design.

\begin{lemma}[Residual geometry and the precision matrix]\label{lem:1}
    Let $D_\Omega=\mathrm{diag}(\Omega)$. Then for every $j$,
    \[
    \alpha_j=\frac{1}{\Omega_{jj}}, \qquad H_{-j}X_j=\frac{X\Omega_{\cdot j}}{\Omega_{jj}}, \qquad \tilde q_j:=\frac{H_{-j}X_j}{\sqrt{\alpha_j}}=\frac{X\Omega_{\cdot j}}{\sqrt{\Omega_{jj}}},
    \]
    and the Gram matrix of the normalised residualised predictors $ Q=[\tilde q_1,\dots,\tilde q_p]\in\mathbb{R}^{n\times p}$ is
    \[ \widetilde\Omega:={Q}^\top Q= D_\Omega^{-1/2}\Omega\, D_\Omega^{-1/2}, \qquad  \widetilde\Omega_{jk}=\frac{\Omega_{jk}}{\sqrt{\Omega_{jj}\Omega_{kk}}},\]
    the precision matrix of the design standardised to unit diagonal. 
\end{lemma}

The radius of $\mathbb{S}_j$ is thus determined by $\Omega_{jj}$ alone, while $\widetilde\Omega$ encodes the geometry of the $p$ residual predictors and hence the extent to which the mirror constructions interact. 

\begin{remark}[Non-degeneracy of the augmented matrix]\label{rem:nondegeneracy}
Since $\mathrm{span}\{X_j^+,X_j^-\}=\mathrm{span}\{X_j,w_j\}$, the augmented matrix $\mathbf{X}^j$ has full column rank $p+1$ if and only if $H_{-j}X_j$ and $w_j$ are linearly independent. Both lie in $\mathrm{col}(H_{-j})$ with norm  $\sqrt{\alpha_j}$, so independence fails only at $w_j=\pm H_{-j}X_j$, where $c_j^+$ or $c_j^-$ vanishes. These two points have probability zero under $v_j\sim\mathrm{Unif}(\mathbb{S}_j)$, so $\mathbf{X}^j$ has full column rank almost surely.
\end{remark}

The $p$ residual subspaces are closely related. Each $\mathrm{col}(H_{-j})$ is the common full-model residual space $\mathrm{col}(H)$ plus a single direction, so the $p$ spheres share an $(n-p)$-dimensional component and differ only in one coordinate. 

\begin{lemma}[Common residual subspace]\label{lem:2}
    For every $j=1,\dots,p$, $H_{-j}=H+\tilde q_j\tilde q_j^\top$, with $\tilde q_j\perp\mathrm{col}(H)$, so that
    \[
    \mathrm{col}(H_{-j})=\mathrm{col}(H)\oplus\mathrm{span}(\tilde q_j),
    \]
    with $\dim\mathrm{col}(H)=n-p=r-1$.
    Writing $\tilde w_j=w_j/\sqrt{\alpha_j}$ for the normalised perturbation and $\gamma_j=\langle\tilde w_j,\tilde q_j\rangle\in[-1,1]$ for its coordinate along the distinguishing direction, if $v_j\sim\mathrm{Unif}(\mathbb{S}_j)$ independently across $j$, then
    \begin{equation}\label{eq:w_decomp}
        \tilde w_j=\sqrt{1-\gamma_j^2}\;\tilde u_j+\gamma_j\,\tilde q_j,
    \end{equation}
    where $\tilde u_j$ is uniform on the unit sphere of $\mathrm{col}(H)$ and independent of $\gamma_j$, and the pairs $(\tilde u_1,\gamma_1),\dots,(\tilde u_p,\gamma_p)$ are mutually independent.
\end{lemma}

Since $\mathrm{col}(H)$ is common to all $j$, one orthonormal basis $U_H\in\mathbb{R}^{n\times(n-p)}$, obtained once from a single factorisation of $X$, serves every variable. Drawing $\xi_j\sim N(0,I_{n-p})$ and $\zeta_j\sim N(0,1)$ mutually independently, 
\begin{equation}\label{eq:w_sampler}
    w_j=\sqrt{\alpha_j}\,\frac{U_H\xi_j+\zeta_j\,\tilde q_j}{\left(\|\xi_j\|^2+\zeta_j^2\right)^{1/2}}, \qquad \gamma_j=\frac{\zeta_j}{\left(\|\xi_j\|^2+\zeta_j^2\right)^{1/2}}, \qquad \tilde u_j=\frac{U_H\xi_j}{\|\xi_j\|}
\end{equation}
realises $w_j$ as $\sqrt{\alpha_j}$ times a uniform point on the unit sphere of $\mathrm{col}(H_{-j})$, since $(U_H,\tilde q_j)$ is an orthonormal basis of that space and $(\xi_j,\zeta_j)$ is standard normal in $\mathbb{R}^{r}$. No per-variable basis $U_j$ is formed. The same identity reduces the $p$ residual norms to a single full-model residual and $p$ inner products,
\begin{equation}\label{eq:hy_decomp}
    \|H_{-j}y\|^2=\|Hy\|^2+(\tilde q_j^\top y)^2.
\end{equation}

Write $a_j^\pm=\alpha_j(1\pm\gamma_j)$. Since $w_j^\top H_{-j}X_j=w_j^\top X_j=\alpha_j\gamma_j$, the residualised mirror columns satisfy 
\begin{equation}\label{eq:c_def}
    c_j^{\pm}=H_{-j}X_j^{\pm}=H_{-j}X_j\pm w_j, \qquad \|c_j^{\pm}\|^2=2a_j^{\pm},
\end{equation}
and by Lemma~\ref{th:1} they are orthogonal, $(c_j^+)^\top c_j^-=0$. The pair $(a_j^+,a_j^-)$ measures the information the two mirror directions carry about $\beta_j^0$, and their minimum, $a_j^\star=\min(a_j^+,a_j^-)=\alpha_j(1-|\gamma_j|)$ governs the analysis throughout.

\subsection{Data Likelihood}
The Bayesian mirror is specified one variable at a time. For each $j$, the working model replaces $X_j$ by its mirror pair, and the mirror statistic for variable $j$ is formed from the resulting coefficients. In contrast to the Gaussian mirror, the $p$ models in the Bayesian mirror require no separate fits of the design, since by Lemma~\ref{lem:2} they share a single factorisation of $X$.

For each variable $j=1,\dots,p$, the working model is
\[y=\mathbf{X}^j\beta^{j}+\epsilon^j, \qquad \epsilon^j\sim N(0,\sigma_j^2I_n),\]
with likelihood
\[p(y\mid\beta^{j},\sigma_j^2,w_j)=(2\pi\sigma_j^2)^{-n/2}\exp\left(-\frac{1}{2\sigma_j^2}\|y-\mathbf{X}^j\beta^{j}\|^2\right).\]
Separating the components that depend on $w_j$ from those that do not,
\begin{equation*} \mathbf{X}^j\beta^{j}=X_j(\beta_j^++\beta_j^-)+X_{-j}\beta_{-j}+w_j(\beta_j^+-\beta_j^-),
\end{equation*}
so the perturbation enters the mean only through the contrast $\beta_j^+-\beta_j^-$, and $X_j$ only through the sum $\beta_j^++\beta_j^-$. Residualising on $X_{-j}$ and using \eqref{eq:c_def} gives the equivalent form,
\begin{equation}\label{eq:resid_decomp}
    H_{-j}\mathbf{X}^j\beta^{j}=c_j^+\beta_j^++c_j^-\beta_j^-.
\end{equation}
Since $(c_j^++c_j^-)/2=H_{-j}X_j$ and $(c_j^+-c_j^-)/2=w_j$, the sum and the difference are the coefficients of the residualised predictor and of the perturbation, respectively. 

\subsection{Prior Specification}\label{sec:prior}
For the mirror coefficients, we assume the conjugate normal--inverse-gamma (NIG) prior of a standard Bayesian linear model,
\[(\beta_j^+,\beta_j^-)^\top\mid\sigma_j^2\sim N\!\left(0,\sigma_j^2\tau^2I_2\right),\qquad \sigma_j^2\sim\mathrm{IG}(a_0,b_0),\]
with $\tau^2$ large and $a_0$, $b_0$ small, so that the prior is diffuse. Since the prior variance of the coefficients is proportional to $\sigma_j^2$, the prior is conjugate to the Gaussian likelihood, so that the posterior of $(\beta_j^+,\beta_j^-,\sigma_j^2)$ given $v_j$ is NIG in closed form. The nuisance coefficients are assigned the flat prior $p(\beta_{-j})\propto1$, independently of $(\beta_j^+,\beta_j^-,\sigma_j^2)$. For the perturbations we take \(v_j\sim\mathrm{Unif}(\mathbb{S}_j)\)
independently across $j$, so that no direction within the orthogonal complement of $\mathrm{col}(X_{-j})$ is preferred to any other, and $v_j$ is independent of $(\beta^j,\sigma_j^2)$.

\subsection{The cut posterior}\label{sec:cut}

Under the full joint posterior, the conditional distribution of $v_j$ depends on $y$ through the contrast $\beta_j^+-\beta_j^-$, so the perturbation would be learned from the response, breaking its exogeneity and, with it, the null symmetry of the mirror statistic.  We therefore adopt the \emph{cut} (modular) distribution \citep{Plummer2015-Cuts,Bayarri2009Modular,Yu2023Variational} as our inferential target, defined as
\begin{equation}\label{eq:cut}
    p(\beta^j,\sigma_j^2,v_j\mid y)=p(v_j)\cdot p(\beta^j,\sigma_j^2\mid v_j,y).
\end{equation}
The cut replaces the conditional $p(v_j\mid y)$ implied by the full posterior with the prior $p(v_j)$, the two coinciding only when the marginal likelihood $p(y\mid v_j)$ is constant on $\mathbb{S}_j$. It blocks the flow of information from $y$ into $v_j$ \citep{Yu2023Variational}, so that $v_j\sim\mathrm{Unif}(\mathbb{S}_j)$ and $w_j$ is independent of $y$. The perturbation is thus auxiliary randomisation that constructs the mirror pair without being learned from the data, as in \citet{Xing2023Controlling-Fal}. Since $v_j$ is not a parameter of the data-generating model, all information that $y$ carries about $\beta^j$ and $\sigma_j^2$ is retained by $p(\beta^j,\sigma_j^2\mid v_j,y)$, the exact posterior of the Bayesian linear model with design $\mathbf{X}^j$. Inference on the coefficients is therefore fully Bayesian, and credible intervals for $\beta_j^++\beta_j^-$ are valid conditional on the perturbation.




Given $v_j$, the design $\mathbf{X}^j$ is fixed and $p(\beta^j,\sigma_j^2\mid v_j,y)$ is the NIG posterior of a Bayesian linear regression. Because the prior on $\beta_{-j}$ is flat, integrating it out leaves the residualised model \eqref{eq:resid_decomp} exactly, and the posterior of $(\beta_j^\pm,\sigma_j^2)$ factorises into a two-dimensional Gaussian for the mirror coefficients and an inverse-gamma for the variance. Writing
\begin{equation}\label{eq:d_kappa}
    d_j^{\pm}=(c_j^{\pm})^\top H_{-j}y=\sqrt{\alpha_j}\,(\tilde q_j^\top y)\pm w_j^\top y, \qquad \kappa_j^{\pm}=\|c_j^{\pm}\|^2+\tau^{-2}=2a_j^{\pm}+\tau^{-2},
\end{equation}
for the projection of the response onto the two mirror directions and the corresponding posterior precision, the coefficient posterior is diagonal,
\begin{equation}\label{eq:beta_reduced}
    \beta_j^{\pm}\mid\sigma_j^2,v_j,y\ \sim\ N\!\left(\frac{d_j^{\pm}}{\kappa_j^{\pm}},\;\frac{\sigma_j^2}{\kappa_j^{\pm}}\right), \qquad \beta_j^{+}\perp\!\!\!\perp\beta_j^{-}\;\bigm|\;\sigma_j^2,v_j,y,
\end{equation}
the independence arising from the orthogonality of Lemma~\ref{th:1}, and the variance posterior is
\begin{equation}\label{eq:sigma_reduced}
    \sigma_j^2\mid v_j,y\ \sim \mathrm{IG}\!\left(a_0+\frac{r}{2},\; b_n\right), \qquad b_n=b_0+\frac{1}{2}\left(\|H_{-j}y\|^2-\frac{(d_j^+)^2}{\kappa_j^+}-\frac{(d_j^-)^2}{\kappa_j^-}\right)
\end{equation}
with $\|H_{-j}y\|^2=\|Hy\|^2+(\tilde q_j^\top y)^2$ by (\ref{eq:hy_decomp}). The coefficient prior thus acts on the mirror construction only by shifting $\kappa_j^\pm$ by $\tau^{-2}$. Equivalently, let $G_j\sim\mathrm{Gamma}(a_0+r/2,1)$ and $z_j^\pm\sim N(0,1)$ be mutually independent and independent of $(\epsilon,w_j)$. A draw from \eqref{eq:beta_reduced}--\eqref{eq:sigma_reduced} is then 
\begin{equation}\label{eq:lem_rand}
    \sigma_j^2=\frac{b_n}{G_j}, \qquad \beta_j^\pm=\frac{d_j^\pm}{\kappa_j^\pm}+\bigg(\frac{\sigma_j^2}{\kappa_j^\pm}\bigg)^{1/2}z_j^\pm.
\end{equation}


\begin{proposition}[Sufficient quantities]\label{prop:suff}
    Let $\hat\beta=\Omega X^\top y$ be the least-squares estimate and suppose $Hy\neq0$. Then $\sqrt{\alpha_j}\,\tilde q_j^\top y=\hat\beta_j/\Omega_{jj}$ and $\|H_{-j}y\|^2=\|Hy\|^2+\hat\beta_j^2/\Omega_{jj}$. Under \eqref{eq:w_sampler}, $h_j=\xi_j^\top U_H^\top y/\|Hy\|\sim N(0,1)$, $\|\xi_j\|^2-h_j^2\sim\chi^2_{n-p-1}$ independently of $h_j$, with $\chi^2_0$ read as zero when $n=p+1$, and 
    \begin{equation}\label{eq:wy}
        w_j^\top y=\frac{\sqrt{\alpha_j}\,\|Hy\|\,h_j+\zeta_j\hat\beta_j/\Omega_{jj}}{\left(\|\xi_j\|^2+\zeta_j^2\right)^{1/2}}.
    \end{equation}
    Consequently, a draw from \eqref{eq:lem_rand} depends on the data only through $\hat\beta_j$, $\Omega_{jj}$ and $\|Hy\|^2$, and requires $O(1)$ operations for each variable and iterate.
\end{proposition}

Proposition~\ref{prop:suff} reduces the computation to a single Cholesky factorisation of $X^\top X$, which yields $\hat\beta$ and $\mathrm{diag}(\Omega)$. Neither the basis $U_H$ nor the perturbations $w_j$ are formed, and at each Monte Carlo iterate $t=1,\dots,T$ and independently for each $j=1,\dots,p$:
\begin{itemize}
    \item Draw $\zeta_j^{(t)},h_j^{(t)}\sim N(0,1)$ and $\|\xi_j^{(t)}\|^2=(h_j^{(t)})^2+\chi^2_{n-p-1}$, and form $\gamma_j^{(t)}$ by \eqref{eq:w_sampler} and $w_j^{(t)\top}y$ by \eqref{eq:wy}, giving $a_j^{\pm(t)}$, $\kappa_j^{\pm(t)}$ and $d_j^{\pm(t)}$.
    \item Draw $G_j^{(t)}$ and $z_j^{\pm(t)}$, and set $\sigma_j^{2(t)}$ and $\beta_j^{\pm(t)}$ by \eqref{eq:lem_rand}.
\end{itemize}
No state is carried between iterates, so each triplet $(v_j^{(t)},\beta^{j(t)},\sigma_j^{2(t)})$ is an independent draw from the cut distribution. The procedure is therefore i.i.d.\ Monte Carlo over perturbation directions.

\subsection{Mirror Statistic}\label{sec:mirror}

\subsubsection{Mirror statistic and null symmetry}
At each Monte Carlo iterate $t$, $t=1,\dots,T$, following \citet{Xing2023Controlling-Fal} and \citet{Dai2023False-Discovery}, the raw mirror statistic for variable $j$ is
\begin{equation*}
    M_j^{(t)}=|\beta_j^{+(t)}+\beta_j^{-(t)}|-|\beta_j^{+(t)}-\beta_j^{-(t)}|.
\end{equation*}
This statistic is not comparably scaled across $j$. By \eqref{eq:beta_reduced} the posterior standard deviation of $\beta_j^{\pm}$ is $\sigma_j/\sqrt{\kappa_j^{\pm}}$, and $\kappa_j^{\pm}=2a_j^{\pm}+\tau^{-2}$ varies with $j$ through $\alpha_j=1/\Omega_{jj}$. For a predictor close to the span of the remaining columns, $\alpha_j$ is small, the likelihood carries little information about its mirror coefficients, and their posterior scale approaches the prior scale $\sigma_j\tau$. Under the null, $M_j^{(t)}$ inherits this scale, so a variable with a diffuse posterior produces a large $|M_j^{(t)}|$ irrespective of whether it is a signal. Since the selection rule compares all $p$ statistics with a single threshold, such variables are over-represented in both tails. We therefore work with the standardised mirror statistic,
\begin{equation*}
    \widetilde M_j^{(t)}=\frac{M_j^{(t)}}{\hat s_j^{(t)}},\qquad  \hat s_j^{(t)}=\sigma_j^{(t)}\theta_j, \qquad \theta_j=\left\{\frac{1}{2}\left(\frac{1}{\kappa_j^+}+\frac{1}{\kappa_j^-}\right)\right\}^{1/2}
\end{equation*}
where $\hat s_j^{(t)}$ is the root-mean-square posterior standard deviation of the pair $(\beta_j^{+(t)},\beta_j^{-(t)})$, and is positive since $\kappa_j^\pm\ge\tau^{-2}>0$. The precisions $\kappa_j^\pm$, and hence $\theta_j$, are non-random given $(X,w_j^{(t)})$, so $\hat s_j^{(t)}$ depends on the response only through the variance draw $\sigma_j^{(t)}$. Since the draws $(v_j^{(t)},\beta^{j(t)},\sigma_j^{2(t)})$ produced by the sampling procedure of Section~\ref{sec:cut} are independent across $t$, the statistics $\{\widetilde M_j^{(t)}\}_{t=1}^{T}$ are, for each $j$, independent and identically distributed replicates conditional on $(X,y)$.

The orthogonality established in Lemma~\ref{th:1}, together with the independence of the perturbation from $y$ under the cut posterior, yields the following exact symmetry result.

\begin{lemma}[Null symmetry of the mirror statistic]\label{lemma1}
    Fix an iterate $t$ and $j\in\mathcal{H}_0$, so that $y=X_{-j}\beta_{-j}^0+\epsilon$, with $\epsilon\sim N(0,\sigma_0^2I_n)$. Let $w_j^{(t)}$ be the realised perturbation at iterate $t$ and $\widetilde M_j^{(t)}$ the standardised mirror statistic evaluated at a draw produced by the sampling procedure of Section~\ref{sec:cut}. Then, for every $s>0$,
    \[\mathbb{P}\!\left(\widetilde M_j^{(t)}\ge s\;\middle|\;w_j^{(t)}\right)=\mathbb{P}\!\left(\widetilde M_j^{(t)}\le -s\;\middle|\;w_j^{(t)}\right).\]
\end{lemma}


\begin{corollary}\label{cor:all_perturbations}
Fix an iterate $t$ and $j\in\mathcal{H}_0$. Then for almost every realisation of the full collection $\mathbf{w}^{(t)}=\{w_j^{(t)}\}_{j=1}^{p}$ and every $s>0$,
\[\mathbb{P}\!\left(\widetilde M_j^{(t)}\ge s\;\middle|\;\mathbf{w}^{(t)}\right)=\mathbb{P}\!\left(\widetilde M_j^{(t)}\le -s\;\middle|\;\mathbf{w}^{(t)}\right).
\]
\end{corollary}

For brevity, we write $\mathbb{P}\!\left(\cdot \; |\mathbf{w}\right)=\mathbb{P}\!_\mathbf{w}(\cdot)$ and $\mathbb{E}\!\left(\cdot \; |\mathbf{w}\right)=\mathbb{E}\!_\mathbf{w}(\cdot)$ throughout.

\subsubsection{Concentration of the mirror counts}\label{sec:toolkit}

Lemma~\ref{lemma1} gives symmetry at each fixed threshold, but the selection threshold is data-driven and depends on all $p$ mirror statistics. Controlling the FDR at that threshold requires the numbers of null and signal statistics exceeding $s$ to concentrate around their means uniformly in $s$. However, at a given iterate, the statistics of different variables are dependent, since they share the noise $\epsilon$, and each involves its own variance draw $\sigma_j^2$, which enters $\widetilde M_j$ non-linearly. We handle the non-linear role of $\sigma_j$ by comparing $\widetilde M_j$ with a surrogate in which $\sigma_j$ is replaced by $\sigma_0$, and the dependence across variables through a spectral measure of dependence among the surrogates.

Fix an iterate $t$ and suppress it. We define the null and signal counts as
\begin{equation*}
V^\pm(s)=\#\{j\in\mathcal{H}_0:\pm\widetilde M_j\ge s\}, \qquad
D^\pm(s)=\#\{j\in\mathcal{H}_1:\pm\widetilde M_j\ge s\}, 
\end{equation*}
so that $R(s)=V^+(s)+D^+(s)$ and $A(s)=V^-(s)+D^-(s)$ are the total counts above $s$ and below $-s$. Write $\bar V^\pm=V^\pm/p_0$, $\bar D^\pm=D^\pm/p_1$, and set
\begin{equation}\label{eq:Gp}
    G_p(s)=\frac{1}{p_0}\sum_{j\in\mathcal{H}_0}\mathbb{P}_\mathbf{w}\!\left(\widetilde M_j\ge s\right),\qquad G_{1,p}^\pm(s)=\frac{1}{p_1}\sum_{j\in\mathcal{H}_1}\mathbb{P}_\mathbf{w}\!\left(\pm\widetilde M_j\ge s\right).
\end{equation}
By Corollary~\ref{cor:all_perturbations}, $G_p(s)$ also equals $p_0^{-1}\sum_{j\in\mathcal{H}_0}\mathbb{P}_\mathbf{w}(\widetilde M_j\le -s)$ for every $s>0$.

The surrogate $\widetilde M_j(\sigma_0)$ replaces $\sigma_j$ by its true value $\sigma_0$ in the coefficient draw \eqref{eq:lem_rand} and in the standardisation,
\begin{equation*}
    \beta_j^\pm(\sigma_0)=\frac{d_j^\pm}{\kappa_j^\pm}+\sigma_0\frac{z_j^\pm}{\sqrt{\kappa_j^\pm}}, \qquad \widetilde M_j(\sigma_0)=\frac{\big|\beta_j^+(\sigma_0)+\beta_j^-(\sigma_0)\big|-\big|\beta_j^+(\sigma_0)-\beta_j^-(\sigma_0)\big|}{\sigma_0\theta_j}, 
\end{equation*}
with counts $V_{\sigma_0}^\pm,D_{\sigma_0}^\pm$. Since $d_j^\pm=a_j^\pm\beta_j^0+(c_j^\pm)^\top\epsilon$, the pair $(\beta_j^+(\sigma_0),\beta_j^-(\sigma_0))$ is affine in the Gaussian vector $(\epsilon,z_j^+,z_j^-)$, and $\sigma_0\theta_j$ is non-random given $(X,\mathbf{w})$. The surrogates are therefore coupled only through the projections $(c_j^\pm)^\top\epsilon$, and their dependence is measured by the largest eigenvalue of the Gram matrix of the normalised residualised mirror columns,
\begin{equation}\label{eq:Chat}
     \Lambda=\lambda_{\max}\!\left(\widetilde C^\top\widetilde C\right), \qquad
    \widetilde C=\Big[\tilde c_1^+,\tilde c_1^-,\dots,\tilde c_p^+, \tilde c_p^-\Big]\in\mathbb{R}^{n\times 2p}, \qquad \tilde c_j^\pm=\frac{c_j^\pm}{\|c_j^\pm\|}.
\end{equation}
$\Lambda$ equals one when the $2p$ directions are mutually orthogonal and $2p$ when they coincide up to sign.

Throughout, $K>0$ is a fixed constant controlling the probability of the exceptional events.
\begin{assumption}[Design regularity]\label{as:design}
There are constants $c_0\in(0,1]$ and $c_v<\infty$, not depending on $p$, such that $\min_j\alpha_j\ge c_0\max_j\alpha_j$ and $\tau^{-2}\le c_v\min_j\alpha_j$.
\end{assumption}

\begin{assumption}[Perturbation regularity]\label{as:spectral}
Conditional on $X$ and $\mathbf{w}$, $\max_{j}|\gamma_j|\le\{2(K+1)\log p/r\}^{1/2}$ and $\Lambda=o\!\left(p_0/\log p_0\right)$.
\end{assumption}

\begin{assumption}[Regime]\label{as:regime}
Assume $\omega_p\big(\psi_{\max}+\sqrt{K\log p}\big)\to0$, where $\psi_j=|\beta_j^{0}|\sqrt{a_j^\star}/\sigma_0$, $\psi_{\max}=\max_j\psi_j$ and
\[\omega_p=\sqrt{\frac{K\log p}{r}}+\frac{1}{r}\left(1+a_0+\frac{b_0}{\sigma_0^2}+K\log p \right)+\frac{\tau^{-2}}{r}\,\frac{\psi_{\max}^2}{\min_j a_j^\star}.\]
\end{assumption}

Assumption~\ref{as:design} concerns the design only, requiring the residual information $\alpha_j=1/\Omega_{jj}$ to be comparable across predictors and the prior term $\tau^{-2}$ not to dominate it. Assumption~\ref{as:spectral} concerns the realised perturbations. Its first part keeps each perturbation away from the degenerate alignments $\gamma_j=\pm1$ of Remark~\ref{rem:nondegeneracy}, and its second part is the dependence condition. Both hold with high probability over the perturbation draw, by Lemma~\ref{lem:4} and Proposition~\ref{prop:spectral} below, so they are conditions on the design rather than on the realised draw. Assumption~\ref{as:regime} is a joint requirement on $n$ and $p$, with its three terms measuring the fluctuation of the residual chi-square, the influence of the prior hyperparameters, and that of the coefficient prior. Under Assumptions~\ref{as:design} and~\ref{as:spectral} with $\psi_{\max}=O(\sqrt{\log p})$, it reduces to $r=n-p+1\gg(\log p)^2$.

\begin{lemma}[Alignment]\label{lem:4}
    For any $K>0$, with probability at least $1-2p^{-K}$ over the perturbation draw,
    \[\max_{j\le p}|\gamma_j|\le\Big\{2(K+1)\log p\,/\,(n-p+1) \Big\}^{1/2}.\]
    Consequently, if $n-p\gg\log p$ then under Assumption~\ref{as:design}, on the same event,
    \begin{equation}\label{eq:balance}
    \min_j\frac{a_j^\star}{\alpha_{\max}}\ge\frac{c_0}{2}, \quad
    \max_j\frac{a_j^{\max}}{a_j^\star}\le1+o(1), \quad \alpha_{\max}=\max_j\alpha_j, \quad a_j^{\max}=\max(a_j^+,a_j^-),
\end{equation}
for all sufficiently large $p$.
\end{lemma}

\begin{lemma}[Coupling]\label{lem:5}
   Under Assumptions~\ref{as:design}--\ref{as:regime}, $\mathbb{P}_\mathbf{w}(\mathcal{A}_p^c)\le C p^{-K}$, where $\mathcal{A}_p=\big\{\max_{j}|\widetilde M_j-\widetilde M_j(\sigma_0)|\le\Delta_p \big\}$ and $\Delta_p=C\omega_p\big(\psi_{\max}+\sqrt{K\log p}\big)\to0$.
\end{lemma}

The surrogate counts concentrate exponentially, and because they and $G_p$ are monotone in $s$, the bound holds uniformly in the threshold. Transferring back to $\widetilde M_j$ through Lemma~\ref{lem:5} gives the following result.

\begin{proposition}[Uniform concentration]\label{prop:conc}
    Under Assumptions~\ref{as:design}--\ref{as:regime}, let $L_{\sigma_0}$ bound the densities of the surrogates $\widetilde M_j(\sigma_0)$, which satisfies $L_{\sigma_0}\le\phi(0)(1+o(1))$ with $\phi$ the standard normal density. Then, for every $\varepsilon>16L_{\sigma_0}\Delta_p$,
    \begin{equation}\label{eq:uniform}
        \mathbb{P}_\mathbf{w}\Big(\sup_{s\in\mathbb{R}}\big|\bar V^\pm(s)-G_p(s)\big|>\varepsilon\Big)\le\frac{3}{\varepsilon}\left\{4e^{-\varepsilon^2p_0/32}+4\exp\!\Big(-\frac{\pi\varepsilon^2p_0}{128\Lambda}\Big)+Cp^{-K}\right\},
    \end{equation}
    and likewise for $\bar D^\pm$ about $G^\pm_{1,p}$ with $p_1$ in place of $p_0$.
\end{proposition}

Taking $\varepsilon\asymp\sqrt{(1\vee\Lambda)\log p_0/p_0}$ in \eqref{eq:uniform} shows $\sup_s|\bar V^\pm-G_p|=o_p(1)$ under Assumption~\ref{as:spectral}. The spectral condition replaces the weak dependence conditions of existing mirror methods. \citet{Dai2023False-Discovery} assume that the variance of the null counts is at most $cp_0^{\alpha}$ for some $\alpha\in(0,2)$, a property of the joint law of the statistics. For linear models, they verify this when the eigenvalues of the covariance matrix of the predictors are bounded away from zero and infinity, the condition also required by \citet{Xing2023Controlling-Fal} in high dimensions. The pointwise bound underlying \eqref{eq:uniform} gives $\mathrm{var}_{\mathbf{w}}(V_{\sigma_0}^\pm(s))\le(4+16\Lambda/\pi)p_0$, so the variance condition holds whenever $\Lambda=O(p_0^{1-\delta})$ for some $\delta>0$. The spectral condition has three advantages. First, $\Lambda$ depends only on $(X,\mathbf{w})$ and involves all predictors rather than the unknown null set, so it can be computed before the response is observed. Second, it depends on the partial rather than the marginal correlations among the predictors. For example, in an equicorrelated setting, the largest eigenvalue of the covariance matrix grows linearly in $p$, whereas the largest eigenvalue of the standardised precision matrix is below two. Third, it yields exponential rather than polynomial tail bounds, uniformly in the threshold. A variance bound gives deviation probabilities that decay only polynomially in $p_0$, which is insufficient for the validity of the aggregated e-values we consider. It remains to certify the spectral part of Assumption~\ref{as:spectral}.

\begin{proposition}[Spectral bound]\label{prop:spectral}
    Let $\widetilde U=[\tilde u_1,\dots, \tilde u_p]$ be the matrix of perturbation directions of Lemma~\ref{lem:2}. Then
    \begin{equation*}
        \sqrt{\Lambda}\le\|Q\|_\mathrm{op}+\|\widetilde U\|_\mathrm{op}, \qquad \| Q\|_\mathrm{op}^2=\lambda_{\max}(\widetilde\Omega)\le\mathrm{cond}(X^\top X),
    \end{equation*}
    where $\mathrm{cond}(\cdot)$ denotes the condition number, and if $n-p\ge16K\log p$ then, with probability at least $1-2p^{1-K}$ given $X$,
    \begin{equation*}
        \|\widetilde U\|_\mathrm{op}\le\sqrt{2}\big(1+\sqrt{p/(n-p)}+\sqrt{2K\log p/(n-p)}\big).
    \end{equation*}
\end{proposition}

The perturbation block contributes $O(1+\sqrt{p/(n-p)})$ for every design, a consequence of the prior rather than an assumption, while the design block is governed by $\lambda_{\max}(\widetilde\Omega)$. The spectral condition thus holds whenever 
\[\Big(\sqrt{\lambda_{\max}(\widetilde\Omega)}+1+\sqrt{p/(n-p)}\Big)^2=o\!\left(\frac{p_0}{\log p_0}\right),\] in particular whenever $\mathrm{cond}(X^\top X)=O(1)$ and $p/(n-p)=o(p_0/\log p_0)$. The condition number is, however, a loose bound. Since $\widetilde\Omega$ is the standardised precision matrix, $\lambda_{\max}(\widetilde\Omega)$ depends on the partial rather than the marginal correlations among the predictors, and it can remain bounded under strong marginal correlation. For the population counterpart of $\widetilde\Omega$, built from $\Sigma^{-1}$, the largest eigenvalue is at most $k$ whenever each row of $\Sigma^{-1}$ has at most $k$ non-zero entries, whatever the correlation level. Block settings with blocks of size $k$ satisfy this, and autoregressive settings, whose precision matrix is tridiagonal, have the largest eigenvalue below two. When $n\gg p$, $\widetilde\Omega$ is close to its population counterpart.

\subsection{Per-iterate FDR control}\label{sec:fdr}
At iterate $t$, the data-driven threshold is,
\begin{equation*}
\tau_q^{(t)}=\inf\left\{s>0:\ \widehat{\mathrm{FDP}}(s)\le q \right\},\quad \mathrm{where} \quad \widehat{\mathrm{FDP}}(s)=\frac{1+A(s)}{R(s)\vee1},
\end{equation*}
with $\inf\emptyset=+\infty$. The selected set is $\widehat{\mathcal{S}}^{(t)}=\{j:\widetilde M_j^{(t)}\ge\tau_q^{(t)}\}$, with realised false discovery proportion $\mathrm{FDP}(\tau_q^{(t)})=V^+(\tau_q^{(t)})/(R(\tau_q^{(t)})\vee1)$.

We impose the following conditions, all as $p\to\infty$ and conditionally on $X$ and $\mathbf{w}$.
\begin{enumerate}
    \item[(C1)] $p_0\to\infty$ and $p_1\to\infty$.
    \item[(C2)] $G_p(s)\to G_0(s)$ and $G^\pm_{1,p}(s)\to G^\pm_1(s)$ for every $s$, the limits not depending on the realisation of $\mathbf{w}$.
    \item[(C3)] There is $s_q>0$, not depending on $p$, with $G^+_1(s_q)>0$ and
    \[\limsup_{p\to\infty}\ \frac{p_0G_0(s_q)+p_1G^-_1(s_q)}{p_0G_0(s_q)+p_1G^+_1(s_q)}\ <\ q.\]
    \item[(C4)] $G_0(s_q)>0$. 
\end{enumerate}

Condition (C2) requires only pointwise convergence of the mean tail functions. Since $\mathbb{E}_\mathbf{w}A(s)=p_0G_p(s)+p_1G^-_{1,p}(s)$ and $\mathbb{E}_{\mathbf{w}}R(s)=p_0G_p(s)+p_1G^+_{1,p}(s)$, the ratio in (C3) is the limit of $\widehat{\mathrm{FDP}}(s_q)$, so (C3) requires the fixed threshold $s_q$ to be eventually feasible. It is the analogue of the condition $\mathrm{FDP}^\infty(t)<q$ of \citet{Xing2023Controlling-Fal}. Condition (C4) keeps the tail mass of the null statistics beyond $s_q$ from vanishing, so that $A$ grows linearly in $p_0$. Together, (C3) and (C4) force $\liminf_p p_1/p_0>0$, since otherwise the ratio tends to one.

The threshold depends on all $p$ statistics, whereas Proposition~\ref{prop:conc} controls the null and signal counts separately, on the scales $p_0$ and $p_1$. Normalising the total counts by $p$ reconciles the two. Let $\pi_0=p_0/p$ and $\pi_1=p_1/p$, and
\begin{equation*}
    \bar R(s)=\pi_0\bar V^+(s)+\pi_1\bar D^+(s), \qquad
    \bar A(s)=\pi_0\bar V^-(s)+\pi_1\bar D^-(s),
\end{equation*}
with conditional means $G_R(s)=\pi_0G_p(s)+\pi_1G^+_{1,p}(s)$ and $G_A(s)=\pi_0G_p(s)+\pi_1G^-_{1,p}(s)$. Since $\pi_0+\pi_1=1$, the uniform deviations of
\begin{equation}\label{eq:etazeta}
    \eta_p^0=\sup_{s\in\mathbb{R}}\max_\pm\big|\bar V^\pm(s)-G_p(s)\big|,\qquad \eta_p^1=\sup_{s\in\mathbb{R}}\max_\pm\big|\bar D^\pm(s)-G^\pm_{1,p}(s)\big|,
\end{equation}
combine into a single bound free of any factor growing in $p$,
\begin{equation}\label{eq:ARdev}
    \sup_{s}\big|\bar R(s)-G_R(s)\big|\le\eta_p, \qquad
    \sup_{s}\big|\bar A(s)-G_A(s)\big|\le\eta_p,
\end{equation}
where $\eta_p=\eta_p^0+\eta_p^1$. By Proposition~\ref{prop:conc}, $\eta_p^0=o_p(1)$ under Assumptions~\ref{as:design}--\ref{as:regime}. The same holds for $\eta_p^1$, since (C1)--(C4) give $p_1\to\infty$ and $\liminf_p p_1/p_0>0$, Assumption~\ref{as:spectral} yields $\Lambda=o(p_1/\log p_1)$. Hence $\eta_p=o_p(1)$. Finally, by Corollary~\ref{cor:all_perturbations},
\begin{equation}\label{eq:diff}
    \sup_{s>0}\big|\bar V^+(s)-\bar V^-(s)\big|\ \le\ 2\eta_p^0.
\end{equation}

\begin{lemma}[Threshold localisation]\label{lem:9}
    Under Assumptions~\ref{as:design}--\ref{as:regime} and conditions (C1)--(C4), there are constants $G_{\min}>0$ and $\varepsilon_0>0$, depending only on $q$ and the limits in (C2)--(C4), such that the event $\mathcal{B}_p=\{\eta_p\le\varepsilon_0\}$ satisfies $\mathbb{P}_\mathbf{w}(\mathcal{B}_p)\to1$ and on $\mathcal{B}_p$, for all sufficiently large $p$,
    \[
    \bar R(s_q)\ \ge\ G_{\min}/2,\qquad \tau_q^{(t)}\ \le\ s_q, \qquad \widehat{\mathrm{FDP}}\big(\tau_q^{(t)}\big)\ \le\ q+\frac{1}{R(\tau_q^{(t)})}.
    \]
\end{lemma}

\begin{theorem}[Per-iterate asymptotic FDR Control]\label{th:2}
    Fix an iterate $t$ and suppose Assumptions~\ref{as:design}--\ref{as:regime} and conditions (C1)--(C4) hold. Then, for any $q\in(0,1)$
    \[\mathrm{FDP}\big(\tau_q^{(t)}\big)\;\le\;q+o_p(1)\quad\text{and}\quad \limsup_{p\to\infty}\;\mathbb{E}_\mathbf{w}\!\left[\frac{\big|\mathcal{H}_0\cap\widehat{\mathcal{S}}^{(t)}\big|}{\big|\widehat{\mathcal{S}}^{(t)}\big|\vee1}\right]\;\le\;q.\]
\end{theorem}

\begin{corollary}\label{cor:ratio}
    Set $c_A=G_0(s_q)/2>0$ and \(\mathcal{D}_p=\mathcal{B}_p\cap\big\{\eta_p^0\le G_0(s_q)/4\big\}.\)
    Then $\mathbb{P}_{\mathbf{w}}(\mathcal{D}_p)\to1$, and on $\mathcal{D}_p$, for all sufficiently large $p$
    \[A\big(\tau_q^{(t)}\big)\;\ge\;c_A\,p_0, \qquad \frac{V^+(\tau_q^{(t)})}{1+A(\tau_q^{(t)})}\;\le\;1+\frac{2\eta_p^0}{c_A}.\]
    In particular
    \(V^+(\tau_q^{(t)})/\left(1+A(\tau_q^{(t)})\right)\;\le\;1+o_p(1)\).
\end{corollary}

Theorem~\ref{th:2} establishes FDR control at each Monte Carlo iterate separately.

\subsection{Global FDR control via the e-Benjamini--Hochberg procedure}\label{sec:ebh}

The FDR control established in Section~\ref{sec:fdr} applies at each iterate separately, producing $T$ selection sets $\widehat{\mathcal{S}}^{(1)},\dots,\widehat{\mathcal{S}}^{(T)}$ that are conditionally independent given $(X,y)$ and must be aggregated into a single rule. Averaging the mirror statistics over iterates and thresholding once does not eliminate the dependence across variables induced by the common response, and the concentration results of Section~\ref{sec:toolkit} would have to be re-established for the averaged statistics.

A different approach, examined by \citet{Ren2023Derandomizing-K} is to apply a majority rule, selecting $j$ when it appears in more than a fraction of the runs. For knockoffs, this rule controls the per-family error rate and the $k$-familywise error rate but not the FDR, and the fraction has no principled data-driven choice. \citet{Ren2023Derandomizing-K} used the conventional choice $0.5$.

The approach we adopt here follows from the recently developed e-value theory. An e-value is a non-negative random variable $e$ satisfying $\mathbb{E}(e)\le1$ under the null. Unlike $p$-values, e-values require no assumption on the dependence among the statistics being combined, so they can be averaged across arbitrarily dependent replicates while preserving validity \citep{Wang2022False-Discovery}. \citet{Ren2024Derandomised-kn} used this to derandomise the knockoff filter, and the same construction applies here with the independent knockoff replicates replaced by the i.i.d.\ Monte Carlo iterates.

To establish FDR control via the e-Benjamini--Hochberg (e-BH) procedure \citep{Wang2022False-Discovery}, it suffices that a relaxed condition holds for the per-iterate e-values, $\sum_{j\in\mathcal{H}_0}\mathbb{E}_{\mathbf{w}}[e_j^{(t)}]\le p$. Following \citet{Ren2024Derandomised-kn}, define the per-iterate e-value
\[e_j^{(t)}=\frac{p\cdot\mathbf{1}\{\widetilde M_j^{(t)}\ge \tau_q^{(t)}\}}{1+\#\big\{k:\widetilde M_k^{(t)}\le -\tau_q^{(t)}\big\}}=\frac{p\cdot\mathbf{1}\{\widetilde M_j^{(t)}\ge\tau_q^{(t)}\}}{1+A(\tau_q^{(t)})}.\]
Summing over the nulls, the relaxed condition is equivalent to
\begin{equation*}
    \mathbb{E}_{\mathbf{w}}\!\left[\frac{V^+(\tau_q^{(t)})}{1+A(\tau_q^{(t)})}    \right]\;\le\;1.
\end{equation*}
Corollary~\ref{cor:ratio} bounds this ratio by $1+o_p(1)$ in probability, but not in expectation. Off the event $\mathcal{D}_p$ the ratio is dominated only by $p_0$, so the expectation also requires  $\mathbb{P}_\mathbf{w}(\mathcal{D}_p^c)=o(p_0^{-1})$, which the exponential bounds of Proposition~\ref{prop:conc} provide.

\begin{lemma}[Tail of the bad event]\label{lem:10}
    Under Assumptions~\ref{as:design}--\ref{as:regime} and conditions (C1)--(C4), with $K\ge2$,
    \[
    p_0\,\mathbb{P}_\mathbf{w}\big(\mathcal{D}_p^c\big)\le C\,p_0\exp\!\left(-\frac{c\,p_0}{1\vee\Lambda}\right)+C\,p_0\,p^{-K}\longrightarrow0,
    \]
    where $c>0$ depends only on $\varepsilon_0$ of Lemma~\ref{lem:9} and $G_0(s_q)$.
\end{lemma}

\begin{theorem}[Asymptotic e-value validity]\label{th:3}
    Under Assumptions~\ref{as:design}--\ref{as:regime} and conditions (C1)--(C4), there is a sequence $\delta_p\to0$, not depending on $j$ or $t$, such that
    \[\sum_{j\in\mathcal{H}_0}\mathbb{E}_{\mathbf{w}}\big[e_j^{(t)}\big]\;\le\;p\,(1+\delta_p),\]
    and consequently, writing $\bar{e}_j=T^{-1}\sum_{t=1}^{T}e_j^{(t)}$,
    \[\sum_{j\in\mathcal{H}_0}\mathbb{E}_{\mathbf{w}}\big[\bar{e}_j\big]=\frac{1}{T}\sum_{t=1}^{T}\sum_{j\in\mathcal{H}_0}\mathbb{E}_{\mathbf{w}}\big[e_j^{(t)}\big]\;\le\;p\,(1+\delta_p), 
    \]
    irrespective of the dependence across iterates or across variables.
\end{theorem}

Applying the e-BH procedure to the averaged e-values at a target level $q^\star\in(0,1)$, the selected set is 
\begin{equation}\label{eq:ebh}
\widehat{\mathcal{S}}=\left\{j:\bar{e}_j\ge\frac{p}{q^\star\hat k}\right\},\qquad \hat k=\max\!\left\{k\in\{1,\dots,p\}:\bar{e}_{(k)}\ge\frac{p}{q^\star k}\right\},
\end{equation}
where $\bar e_{(1)}\ge\dots\ge\bar e_{(p)}$ are the order statistics and $\widehat{\mathcal{S}}=\emptyset$ if no such $k$ exists. Theorem~\ref{th:3} and the e-BH guarantee give global FDR control.

\begin{theorem}[Global FDR control]\label{th:4}
Under Assumptions~\ref{as:design}--\ref{as:regime} and conditions (C1)--(C4), let $\widehat{\mathcal{S}}$ be given by \eqref{eq:ebh}. Then, conditionally on $X$ and $\mathbf{w}$,
    \[\mathrm{FDR}\big(\widehat{\mathcal{S}}\big)=\mathbb{E}_{\mathbf{w}}\!\left[\frac{\big|\mathcal{H}_0\cap\widehat{\mathcal{S}}\big|}{\big|\widehat{\mathcal{S}}\big|\vee1}\right]\;\le\;q^\star\big(1+\delta_p\big)=q^\star\big(1+o(1)\big),\]
    irrespective of the dependence across iterates or across variables. 
\end{theorem}

\begin{corollary}[Unconditional control]\label{cor:uncond}
    Suppose in addition that $\varepsilon_0$, $c$ and $\Lambda$ in Lemma~\ref{lem:10} admit bounds not depending on the realisation of $\mathbf{w}$, as is the case, by Lemma~\ref{lem:4} and Proposition~\ref{prop:spectral}, on an event of probability at least $1-4p^{1-K}$. Then $\delta_p$ admits a bound $\bar\delta_p\to0$ free of $\mathbf{w}$, and 
    \[
    \mathbb{E}\!\left[\frac{\big|\mathcal{H}_0\cap\widehat{\mathcal{S}}\big|}{\big|\widehat{\mathcal{S}}\big|\vee1}\right]\le q^\star\big(1+\bar\delta_p\big)+4p^{1-K}=q^\star\big(1+o(1)\big).
    \]
\end{corollary}

\subsection{High-Dimensional Bayesian Mirror}\label{sec:high_dim}
The construction of the mirror variables requires $r=n-p+1\ge2$, that is $n>p$. When $p\ge n$, the orthogonal complement of $\mathrm{col}(X_{-j})$ contains no sphere on which to place the perturbation. The same obstacle is met by \citet{Xing2023Controlling-Fal} and \citet{Dai2023False-Discovery}, and is resolved in the same way, by screening to a feasible dimension before applying the mirror construction.

We use a two-stage procedure on split data. The observations are partitioned into independent subsets $(y^{(1)},X^{(1)})\in\mathbb{R}^{n_1}\times\mathbb{R}^{n_1\times p}$ and $(y^{(2)},X^{(2)})\in\mathbb{R}^{n_2}\times \mathbb{R}^{n_2\times p}$, with $n_1+n_2=n$, each satisfying
\[
y^{(k)} = X^{(k)}\beta^0 + \epsilon^{(k)}, \quad \epsilon^{(k)} \sim N(0, \sigma_0^2 I_{n_k}), \quad k = 1, 2,
\]
the first used for lasso screening and the second for Bayesian mirror inference. If the same observations were used for both stages, the screened set $\widehat{J}$ would be a function of the response entering the second stage, and the noise would be conditioned on the screening event. This is a stronger separation than in \citet{Xing2023Controlling-Fal}, who screen and infer on the same data and correct for the selection event through post-selection inference, and than in \cite{Dai2023False-Discovery}, whose mirror statistic incorporates estimates from the screening stage.

Screening is subject to two errors. The first, that is a null spuriously retained in $\widehat{J}$ and then passed to the second stage, is handled by the FDR-controlled selection of the second stage, at a cost only in the dimension of the reduced problem. The second, that is a signal omitted from $\widehat{J}$ is more problematic. Specifically, it misspecifies the second-stage regression and destroys the null symmetry of Lemma~\ref{lemma1} for the retained nulls correlated with it, so that, besides the loss of power, it inflates the FDR. The screen should therefore favour retaining too many variables over too few.

\paragraph{Stage 1: Lasso screening.}
The first subset is used to solve the lasso problem over all $p$ variables,
\[
\tilde{\beta} = \arg\min_{\beta \in \mathbb{R}^p} \left\{\frac{1}{n_1} \|y^{(1)} - X^{(1)}\beta\|^2 + \lambda_{n_1} \|\beta\|_1 \right\},
\]
with regularisation parameter chosen as
\[
\lambda_{n_1} = \delta\hat{\sigma}\sqrt{\frac{\log p}{n_1}},
\]
where $\hat\sigma$ is estimated on the screening subset. The screened set and its cardinality are $\widehat{J}=\{j:\tilde\beta_j\neq0\}$ and $\hat{p}=|\widehat{J}|$. Under the compatibility and signal-strength conditions of Theorem~6.1 of \citet{Buhlmann2011-Stat}, the choice $\delta=4$ yields the sure screening property  
\begin{equation*}
\mathbb{P}(\mathcal{H}_1\subseteq\widehat{J})\geq 1-2/p,
\end{equation*}
with \citet{Xing2023Controlling-Fal} adopting the same calibration for the pre-screening step of the Gaussian mirror. 

\paragraph{Stage 2: Bayesian Mirror inference.}
The second subset is restricted to the screened features, forming the reduced design $X_{\widehat{J}}^{(2)}\in\mathbb{R}^{n_2\times\hat{p}}$. All subsequent inference proceeds exclusively on $(y^{(2)},X_{\widehat{J}}^{(2)})$ exactly as in Sections~\ref{sec:bm_variables}--\ref{sec:ebh}, with all quantities recomputed from the screened predictors. In particular, $\hat r=n_2-\hat p+1$ and $\widehat\Lambda=\lambda_{\max}(\widetilde C^\top\widetilde C)$ for the matrix of normalised residualised mirror columns of $X^{(2)}_{\widehat{J}}$, and
\[
\hat p_0=\big|\mathcal{H}_0\cap\widehat{J}\big|,\qquad 
\hat p_1=\big|\mathcal{H}_1\cap\widehat{J}\big|,\qquad \hat p_0+\hat p_1=\hat p.
\]
Variables in $\{1,\dots,p\}\setminus \widehat{J}$ are assigned $\widetilde M_j=0$ and are never selected.

\paragraph{Validity of the split procedure.}
Because $\widehat{J}=\widehat{J}(X^{(1)},y^{(1)})$ is a function of the screening subset alone and $(X^{(2)},y^{(2)})$ is independent of it, conditioning on $\widehat{J}$ leaves the distribution of $\epsilon^{(2)}$ unchanged so $w_j\perp\!\!\!\perp y^{(2)}\mid X_{\widehat{J}}^{(2)},\widehat{J}$ exactly. Lemma~\ref{lemma1} therefore applies to every null variable in $\widehat{J}$ without modification and without any screening correction. 

\begin{proposition}[FDR control after screening]\label{prop:hd}
    Let $\mathcal{F}_p$ denote the event on which all of the following hold:
    \begin{enumerate}
        \item[(S1)] $\mathcal{H}_1\subseteq\widehat{J}$ (sure screening),
        \item[(S2)] $X^{(2)}_{\widehat{J}}$ has full column rank and $\hat p<n_2$,
        \item[(S3)] Assumptions~\ref{as:design}--\ref{as:regime} and conditions (C1)--(C4) hold for the screened problem with $n$, $p$, $r$, $\Lambda$, $\mathcal{H}_0$, $p_0$, $p_1$ replaced by $n_2,\hat p,\hat r, \widehat\Lambda,\mathcal{H}_0\cap\widehat{J},\hat p_0,\hat p_1$, with all constants not depending on the realisation of $\widehat{J}$ and $\mathbf{w}$, as in Corollary~\ref{cor:uncond}.
    \end{enumerate}
    If $\mathbb{P}(\mathcal{F}_p)\to1$, then the conclusions of Theorem~\ref{th:2} and Corollary~\ref{cor:uncond} hold for the two-stage procedure,
    \[
    \mathrm{FDP}\big(\tau_q^{(t)}\big)\le q+o_p(1),\qquad
    \limsup_{p\to\infty}\ \mathbb{E}\!\left[\frac{\big|\mathcal{H_0}\cap\widehat{\mathcal{S}}\big|}{\big|\widehat{\mathcal{S}}\big|\vee1}\right]\le\ q^\star,
    \]
    where $\widehat{\mathcal{S}}$ is the set returned by \eqref{eq:ebh}.
\end{proposition}

\begin{remark}[Sure screening]\label{rem:sure-screening}
    Condition (S1) cannot be dropped in general. If $\mathcal{M}=\mathcal{H}_1\setminus\widehat{J}$ is non-empty, then for a retained null $j\in\mathcal{H}_0\cap\widehat{J}$ the omitted term $X^{(2)}_\mathcal{M}\beta^0_{\mathcal{M}}$ enters the projected response, so $ d_j^\pm=(c_j^\pm)^\top X^{(2)}_{\mathcal{M}}\beta^0_{\mathcal{M}}+(c_j^\pm)^\top\epsilon^{(2)}$ is no longer centred and the null symmetry of Lemma~\ref{lemma1} fails. A retained null correlated with an omitted signal then has an asymmetric statistic with generically non-zero mean, so omission inflates the FDR rather than just reducing power. Sure screening therefore ensures that the second-stage regression is correctly specified.
\end{remark}

With a single partition, the selected set depends on which observations enter the screening subset, and a signal omitted by the screen is lost to the second stage. For this reason, we recommend, in practice, that the two-stage procedure is repeated over $L$ independent partitions, which we refer to as the multi-split Bayesian mirror. Each partition yields its own screened set and, from its inference subset, e-values for the retained variables, where a zero e-value is assigned to the screened out variables, and the retained e-values are rescaled by the ratio of $p$ to the size of that partition's screened set, so that each partition contributes a valid set of e-values for all $p$ hypotheses. These are averaged across partitions and the e-BH procedure is applied once to the averages. Averaging removes the dependence of the selection on a particular partition, and a signal missed by one screen retains evidence from the partitions in which it is screened. Since averages of e-values remain valid under arbitrary dependence between partitions, the FDR guarantee of Proposition~\ref{prop:hd} extends to the multi-split procedure whenever $n_2\,\mathbb{P}(\mathcal{F}_p^c)\to0$, with the exceptional perturbation event of Corollary~\ref{cor:uncond} included in $\mathcal{F}_p^c$. The complete procedure is summarised in Algorithm~\ref{alg:bm}, and since it requires only $n_2>\hat p$, it applies equally when $n>p$.

\begin{algorithm}[htbp]
\caption{Multi-split Bayesian mirror}
\label{alg:bm}
\begin{algorithmic}[1]
\Require $(y, X)$, $n_1$, $L$, $T$, $(q,q^\star)$ satisfying \eqref{eq:alpha-condition}, $(\tau^2,a_0,b_0)$
\For{$l = 1, \dots, L$}
    \State Split $\{1,\dots,n\}$ at random into $(y^{(1)}, X^{(1)})$ and $(y^{(2)}, X^{(2)})$ of sizes $n_1$ and $n_2=n-n_1$
    \State$ \widehat{J}_l \gets \{j : \tilde\beta_j \neq 0\}$, with $\tilde\beta$ the lasso fit on $(y^{(1)}, X^{(1)})$ \Comment{Stage 1}
    \State Compute $\hat\beta$, $\mathrm{diag}(\Omega)$ and $\|Hy^{(2)}\|^2$ from $(y^{(2)}, X^{(2)})_{\widehat{J}_l}$ \Comment{Stage 2} 
    \For{$t = 1, \dots, \lfloor T/L \rfloor$}
        \For{$j\in\widehat J_l$} 
           \State Draw $\gamma_j$ and $w_j^\top y^{(2)}$ by \eqref{eq:w_sampler} and \eqref{eq:wy}
           \State Draw $\sigma_j^2$ and $\beta_j^\pm$ by \eqref{eq:lem_rand}, and form $\widetilde M_j^{(t)}$
        \EndFor
        \State $\tau_q^{(t)}\gets$ threshold on $\{\widetilde M_j^{(t)}\}_{j\in\widehat{J}_l}$, \quad $A^{(t)} \gets \#\{j : \widetilde M_j^{(t)} \le -\tau_q^{(t)}\}$
    \EndFor
    \State $e_j^{(l)} \gets \dfrac{p}{\lfloor T/L \rfloor}\displaystyle\sum_{t=1}^{\lfloor T/L \rfloor} \frac{\mathbf{1}\{\widetilde M_j^{(t)} \ge \tau_q^{(t)}\}}{1 + A^{(t)}}$ for $j \in \widehat{J}_l$, and $e_j^{(l)} \gets 0$ otherwise
\EndFor
\State $\bar e_j \gets L^{-1}\sum_{l=1}^L e_j^{(l)}$ \quad $j = 1, \dots, p$
\State $\widehat{\mathcal{S}}\gets$ e-BH at level $q^\star$ on $(\bar e_1, \dots, \bar e_p)$ as in \eqref{eq:ebh}.
\Ensure $\widehat{\mathcal{S}}$.
\end{algorithmic}
\end{algorithm}

\subsection{Power Analysis}
The properties of the mirror statistic under the null hypothesis, together with FDR control at the per-iterate level and its aggregation via e-values have been established. We now turn to its behaviour under the alternative.

\subsubsection{Per-iterate Power}\label{sec:per-iterate-power}
Fix an iterate $t$ and $j\in\mathcal{H}_1$, suppressing the superscript. The mirror statistic separates into a deterministic term carrying the signal and a mean-zero fluctuation, and the power analysis rests on bounding the two apart. Under $y=X\beta^0+\epsilon$ the projected response is $H_{-j}y=\beta_j^0H_{-j}X_j+H_{-j}\epsilon$, so by \eqref{eq:c_def} and \eqref{eq:d_kappa}
\begin{equation}\label{eq:d-comp}
    d_j^{\pm}= a_j^{\pm}\beta^0_j+g_j^{\pm}, \qquad g_j^{\pm}:=(c_j^{\pm})^\top\epsilon\sim N\big(0,2\sigma_0^2a_j^\pm\big),
\end{equation}
where $g_j^+\perp\!\!\!\perp g_j^-$ by Lemma~\ref{th:1}. Substituting into \eqref{eq:lem_rand} and writing $f(a)=a/(2a+\tau^{-2})$, so that $a_j^\pm/\kappa_j^\pm=f(a_j^\pm)$, the two halves of the mirror statistic are
\[\underbrace{\beta_j^++\beta_j^-}_{u_j}=\beta_j^0\lambda_u+E_u, \qquad
   \underbrace{\beta_j^+-\beta_j^-}_{\Delta_j}=\beta_j^0\lambda_{\Delta}+E_{\Delta},
\]
with signal loadings $\lambda_u = f(a_j^+) + f(a_j^-)$ and $\lambda_{\Delta} = f(a_j^+)-f(a_j^-)$, and noise terms
\[E_u = \frac{g_j^+}{\kappa_j^+} + \frac{g_j^-}{\kappa_j^-}+\sigma_j\!\left(\frac{z_j^+}{\sqrt{\kappa_j^+}} +\frac{z_j^-}{\sqrt{\kappa_j^-}}\right),\qquad  E_\Delta = \frac{g_j^+}{\kappa_j^+} - \frac{g_j^-}{\kappa_j^-}+\sigma_j\!\left(\frac{z_j^+}{\sqrt{\kappa_j^+}} -\frac{z_j^-}{\sqrt{\kappa_j^-}}\right). \]
Since $a_j^+-a_j^-=2\alpha_j\gamma_j$, the contrast loading is $\lambda_\Delta=2\alpha_j\gamma_j\tau^{-2}/(\kappa_j^+\kappa_j^-)$, which vanishes in the flat prior limit $\tau^{-2}\to0$, where the contrast $\Delta_j$ carries no signal. 

\begin{lemma}[Deterministic signal separation]\label{lem:power-A}
    For every $w_j$ with $a_j^{\star}>0$ and every $\tau^{-2}\ge0$,
    \begin{equation*}
        \lambda_u-|\lambda_{\Delta}|=2f(a_j^{\star})=\frac{2a_j^\star}{2a_j^\star+\tau^{-2}}>0, 
    \end{equation*}
    and pointwise, 
    \begin{equation*}
        M_j\ge2f(a_j^\star)\,|\beta_j^0|-|E_u|-|E_\Delta|. 
    \end{equation*}
\end{lemma}

Selection uses the standardised mirror statistic, $\widetilde M_j^{(t)}$, so the separation must be expressed on that scale. Dividing the bound of Lemma~\ref{lem:power-A} by $\hat s_j=\sigma_j\theta_j$, with $\psi_j$ the standardised signal strength of Assumption~\ref{as:regime}, and controlling $\sigma_0/\sigma_j$ through Lemma~\ref{lem:5} gives the following.

\begin{lemma}[Separation on the standardised signal]\label{lem:power-std}
    Under Assumptions~\ref{as:design}--\ref{as:regime}, with $\mathbb{P}_\mathbf{w}$-probability at least $1-Cp^{-K}$, for every $j\in\mathcal H_1$,
    \begin{equation}\label{eq:std-sep}
        \widetilde M_j\ge c_\psi\psi_j\big(1-C\omega_p\big)-\Xi_j,\qquad \Xi_j:=\frac{|E_u|+|E_\Delta|}{\sigma_j\theta_j},\qquad
        c_\psi=\sqrt{\frac{2}{1+c_v}}\in(0,\sqrt{2}],
    \end{equation}
    with $c_\psi\to\sqrt2$ as $\tau^{-2}\to0$.
\end{lemma}

The factor $1-C\omega_p$ carries the rate $\omega_p$ of Assumption~\ref{as:regime}, so the loss $C\omega_p c_\psi\psi_j$ is negligible. On the standardised scale, the residual $\Xi_j$ is bounded by a constant multiple of the largest of four standard normal variables, hence the signal must clear the threshold by a margin exceeding this fluctuation.

\begin{itemize}
    \item[(A1)] \emph{Beta-min.} There is a sequence $m_p$, bounded below by a positive constant, with $c_\psi\psi_j\ge s_q+m_p$ for all sufficiently large $p$, where $s_q$ is the threshold of (C3). 
\end{itemize}

For designs with independent rows, $\alpha_j$ is of order $n-p$, so (A1) amounts to $|\beta_j^0|\gtrsim\sigma_0(s_q+m_p)/\sqrt{n-p}$, the classical beta-min of order $\sigma_0\sqrt{\log p/(n-p)}$ when $m_p\asymp\sqrt{\log p}$. 

\begin{theorem}[Per-iterate signal recovery]\label{th:5}
    Fix an iterate $t$ and $j\in \mathcal{H}_1$. Under Assumptions~\ref{as:design}--\ref{as:regime}, conditions (C1)--(C4) and (A1), there are $C,C_1>0$, such that for all sufficiently large $p$,
    \[\mathbb{P}_\mathbf{w}\!\bigl(j\in\widehat{S}^{(t)}\bigr)\ge 1-8\exp\!\Big(-\frac{m_p^2}{8C_1^2}\Big)-Cp^{-K}-\mathbb{P}_\mathbf{w}\big(\mathcal{B}_p^c\big),
   \]
   where $\mathcal{B}_p$ is the event of Lemma~\ref{lem:9}. In particular, the right-hand side tends to one whenever $m_p\to\infty$.
\end{theorem}

\begin{corollary}[Per-iterate global power]\label{cor:power-global}
    Suppose (A1) holds uniformly over $j\in \mathcal{H}_1$, with 
    \(m_p\ge 2C_1\sqrt{2(K+1)\log p_1}\). Then, conditionally on $(X,\mathbf{w})$,
    \begin{equation*}
    \mathbb{P}_\mathbf{w}\big(\mathcal{H}_1\subseteq\widehat{\mathcal{S}}^{(t)}\big)\longrightarrow1, \qquad \Pi^{(t)}=\mathbb{E}_\mathbf{w}\!\left[\frac{\bigl|\widehat{S}^{(t)}\cap \mathcal{H}_1\bigr|}{p_1}\right]\longrightarrow1. 
    \end{equation*}
\end{corollary}

The quantity $a_j^\star$ acts as the effective sample size for $\beta_j^0$ within the mirror construction, and $\psi_j$ as the corresponding signal-to-noise ratio. The prior enters only through $\tau^{-2}$. As $\tau^{-2}\to0$, the separation constant $2f(a_j^\star)$ rises to one, $c_\psi$ to $\sqrt2$, and the signal lost to the contrast, $\lambda_\Delta\propto\tau^{-2}$, vanishes. The resulting beta-min is of order $\sigma_0\sqrt{\log p/(n-p)}$, with no dependence on $p_1$. By contrast, \citet{Xing2023Controlling-Fal} do not provide any power guarantees, while \citet{Dai2023False-Discovery} obtain $\liminf\mathrm{Power}=1$ under a signal-strength condition of order $\sqrt{p_1\log p/n}$.

\subsubsection{Power of the aggregated procedure}\label{sec:ebh-power}
Theorem~\ref{th:5} and Corollary~\ref{cor:power-global} concern the selection set at a single Monte Carlo iterate. We now show that the aggregated e-BH procedure recovers the entire active set with probability tending to one, under the per-iterate conditions already imposed and a relation between the two levels $q$ and $q^\star$.

\begin{theorem}[Power of the e-BH aggregated procedure]\label{th:6}
    Suppose Assumptions~\ref{as:design}--\ref{as:regime} and conditions (C1)--(C4) hold, that (A1) holds uniformly over $j\in\mathcal{H}_1$ with $m_p\ge2C_1\sqrt{2(K+1)\log p_1}$, and that the e-BH level satisfies
    \begin{equation}\label{eq:alpha-condition}
        q^\star>\frac{q}{1-q}.
    \end{equation}
    Suppose further that the number of iterates $T$, which may grow with $p$, satisfies, for every $c>0$,
    \begin{equation}\label{eq:T-condition}
        T\Big\{p_1^{-K}+ \exp\Big(-\frac{c\,p_0}{1\vee\Lambda}\Big)\Big\}\longrightarrow0.
    \end{equation}
    Then, conditionally on $X$ and $\mathbf{w}$,
    \begin{equation*}
        \mathbb{P}_{\mathbf{w}}\bigl(\mathcal{H}_1\subseteq\widehat{\mathcal{S}}\bigr) \longrightarrow1, \qquad      \Pi^{\text{e-BH}}:=\mathbb{E}_\mathbf{w}\!\left[\frac{\bigl|\widehat{\mathcal{S}}\cap \mathcal{H}_1\bigr|}{p_1} \right]\longrightarrow1.
    \end{equation*}
\end{theorem}

Condition \eqref{eq:T-condition} holds for any fixed $T$, by Assumption~\ref{as:spectral}, and for $T=o(p_1^{K})$ whenever $p_0/(1\vee\Lambda)\gg\log p_1$. The condition $q^\star>q/(1-q)$, equivalently $q<q^\star/(1+q^\star)$, links the per-iterate level $q$ to the aggregation level $q^\star$. When every signal is selected at an iterate, $R(\tau_q^{(t)})=p_1+V^+(\tau_q^{(t)})$, and together with Corollary~\ref{cor:ratio} this bounds each signal's e-value below by $p(1-q)/(qp_1)$ asymptotically. Condition \eqref{eq:alpha-condition} is the point at which this bound exceeds the e-BH threshold $p/(q^\star p_1)$, and when it fails the argument no longer guarantees threshold crossing. \citet{Ren2024Derandomised-kn} derive the same boundary heuristically for derandomised knockoffs, in a scenario where every run selects all signals, and Theorem~\ref{th:6} makes this argument rigorous, with no sparsity condition beyond (C3) and (C4) and no signal-strength assumption beyond (A1). Within the admissible range, a smaller $q$ enlarges the e-value bound but raises $s_q$ and hence tightens (A1). \citet{Ren2024Derandomised-kn} find empirically that $q=q^\star/2$ works well, and this choice satisfies \eqref{eq:alpha-condition} for every $q^\star\in(0,1)$.

\begin{remark}
    Theorem~\ref{th:6} is conditional on $(X,\mathbf{w})$. By Lemma~\ref{lem:4} and Proposition~\ref{prop:spectral}, Assumption~\ref{as:spectral} holds with probability $1-4p^{1-K}$ given $X$, and on that event $a_j^\star=\alpha_j(1+o(1))$, so Assumption~\ref{as:regime} and (A1) reduce to conditions on $X$ and $\beta^0$. Hence $\mathbb{P}(\mathcal{H}_1\subseteq\widehat{\mathcal{S}}\mid X)\to1$ by the tower property. The FDR guarantee of Theorem~\ref{th:4} holds for every $T$, whereas the power statement requires \eqref{eq:T-condition}, since it controls all iterates simultaneously.
\end{remark}

\subsubsection{Power in the high-dimensional setting}\label{sec:hd-power}

The power statements transfer to the two-stage procedure under the event $\mathcal{F}_p$ that carries the FDR guarantee. On $\mathcal{F}_p$ the screened model is correctly specified, and every quantity of Sections~\ref{sec:per-iterate-power} and~\ref{sec:ebh-power} is computed on the screened predictors $X^{(2)}_{\widehat{J}}$, and denoted with a hat. In particular, $\hat r=n_2-\hat p +1$ is the residual dimension, $\hat\psi_j=|\beta_j^0|\sqrt{\hat a_j^{\star}}/\sigma_0$ the standardised signal strength, and $\hat p_1=|\mathcal{H}_1\cap\widehat{J}|$, which equals $p_1$ under (S1).

\begin{proposition}[Power on the screened problem]\label{prop:hd-power}
    Let $\mathcal{F}_p$ be the event of Proposition~\ref{prop:hd}, with $\mathbb{P}(\mathcal{F}_p)\to1$. Suppose that on $\mathcal{F}_p$ condition (A1) holds uniformly over $j\in\mathcal{H}_1$ for the screened problem, with $\hat\psi_j$ in place of $\psi_j$ and $m_p\ge2C_1\{2(K+1)\log p_1\}^{1/2}$, and that \eqref{eq:alpha-condition} and \eqref{eq:T-condition} hold for the screened problem. Then
    \begin{equation}\label{eq:hd-sure-discovery}
    \mathbb{P}\bigl(\mathcal{H}_1\subseteq\widehat{\mathcal{S}}\bigr)\longrightarrow1,\qquad
        \mathbb{E}\left[\frac{\big|\widehat{\mathcal{S}}\cap\mathcal{H}_1\big|}{p_1}\right]\longrightarrow1.
    \end{equation}
\end{proposition}

For designs with independent rows, $\hat\alpha_j$ is of order $n_2-\hat p$, so (A1) is a beta-min of order $\sigma_0\sqrt{\log p_1/(n_2-\hat p)}$. The same conclusions hold for the multi-split procedure whenever $L\,\mathbb{P}(\mathcal{F}_p^c)\to0$, since on the event that every partition satisfies $\mathcal{F}_p$, the rescaled e-value of each signal in each partition obeys the same lower bound, and so does their average.

\begin{remark}
    Conditions (C3) and (C4) require $\liminf\hat p_1/\hat p_0>0$, which in the low-dimensional setting of Section~\ref{sec:fdr} confines the theory to non-sparse settings with $p_1\gtrsim p_0$. Here the requirement is delivered by the screening stage rather than assumed. By (S1), $\hat p_1=p_1$, so if $\hat p\le Cp_1$, then $\hat p_0=\hat p-p_1\le (C-1)\hat p_1$. Provided $\hat p_0\to\infty$, the two-stage procedure recovers the full active set however sparse it is, as long as the screened set remains proportional to $\mathcal{H}_1$.
\end{remark}

\section{Experimental Results}
We compare the Bayesian mirror against model-X knockoffs \citep{Candes2018Panning-for-Gol}, derandomised knockoffs aggregated by e-BH \citep{Ren2024Derandomised-kn}, Gaussian mirror \citep{Xing2023Controlling-Fal}, data splitting and multiple data splitting \citep{Dai2023False-Discovery}, and nullstrap \citep{wang2025nullstrapsimplehighpowerfast}. The target level is $q^\star=0.1$ throughout. The Bayesian mirror is implemented using the multi-split procedure of Algorithm~\ref{alg:bm} with $T=1000$ Monte Carlo iterates, drawn independently, and aggregated by the e-BH procedure at level $q^\star$. Throughout, the prior hyperparameters are $\tau^2=n_2$, $a_0=0.01$ and $b_0=0.01$. Derandomised knockoffs use $50$ independent knockoff draws aggregated at the same level, with individual runs calibrated at $q=0.03$, which gains power over $q=0.05$. For both knockoff procedures, the covariance matrix is estimated by shrinkage towards the identity, the default of the \texttt{knockoff} R package, and the knockoffs are constructed by the approximate semidefinite program. Multiple data splitting uses $50$ independent splits. Nullstrap uses the lasso with the penalty set to half of the cross-validated minimiser and calibrates its correction factor on five positive controls, as in its R implementation. For each method, we report the empirical FDR, 
which is the realised discovery proportion averaged across replications, the empirical power, which is the average proportion of the 
true signals recovered, and, as a measure of stability, the false discovery exceedance, defined as the proportion of replications whose FDP exceeds $q^\star$. All methods were implemented in Python after porting their respective R software implementations. 
Data generating mechanisms as well as software implementations of all methods and experiments are publicly available and reproducible at \url{https://github.com/theopapazoglou/Bayesian-Mirror.git}. 

\subsection{Simulation Studies}\label{sec:sim}

We fix $n=800$ and $p=6000$, and the rows of the design matrix are drawn independently from $N(0,\Sigma)$. The support $\mathcal H_1$ is a uniformly drawn subset of size $p_1=70$, the coefficients 
are drawn as $\beta_j^0\sim N(0,1)$, and the response follows the linear model, $y=X\beta^0+\epsilon$, with $\epsilon\sim N(0,I_n)$. We consider three dependence structures. 
\begin{itemize}
\item 
Autoregressive Toeplitz covariance: $\Sigma_{ij}=\rho^{|i-j|}$,  with $\rho\in\{0,0.2,0.4,0.6,0.8\}$. Correlation decays with the distance between predictors. 

\item Block correlation structure: $\Sigma_{ij}=\rho$ for $i\neq j$ in the same block and $\Sigma_{ij}=0$ across blocks, with $\rho\in\{0.3,0.6,0.9\}$. The predictors are partitioned into consecutive blocks of $10$. Under this setting every predictor is equicorrelated with the other members of its block, and a signal may share its block with other signals or with null predictors. 
\item 
Proxy twins setting: it isolates null predictors that act as proxies for the signal. Each signal $X_j$, $j\in\mathcal H_1$, is accompanied by $3$ null twins, 
\[
X_{j,l}=rX_j+\sqrt{1-r^2}\,Z_{j,l},\quad l=1,2,3, \quad r\in\{0.4,0.6,0.8\},
\]
where $Z_{j,l}$ are independent standard normal vectors, and the remaining $p-4p_1$ null predictors are independent standard normal. Each twin has unit variance and correlation $r$ with its own signal, two twins of the same signal have correlation $r^2$, and twins are uncorrelated with all other predictors. The signals and their twins are placed at uniformly drawn positions.
Since every correlated null predictor is attached to a single signal and carries no effect, any selected twin is a false discovery induced by its signal.
\end{itemize}

For the multi-split Bayesian mirror of Algorithm~\ref{alg:bm}, we use $L=20$ partitions, a screening fraction of $n_1/n=0.8$, a per-iterate level of $q=0.03$ and $T=1000$ Monte Carlo iterates. All methods are applied to the same data sets, and each setting is replicated $100$ times.

\paragraph{Autoregressive correlation.}

\begin{figure}
    \centering
    \includegraphics[width=1\linewidth]{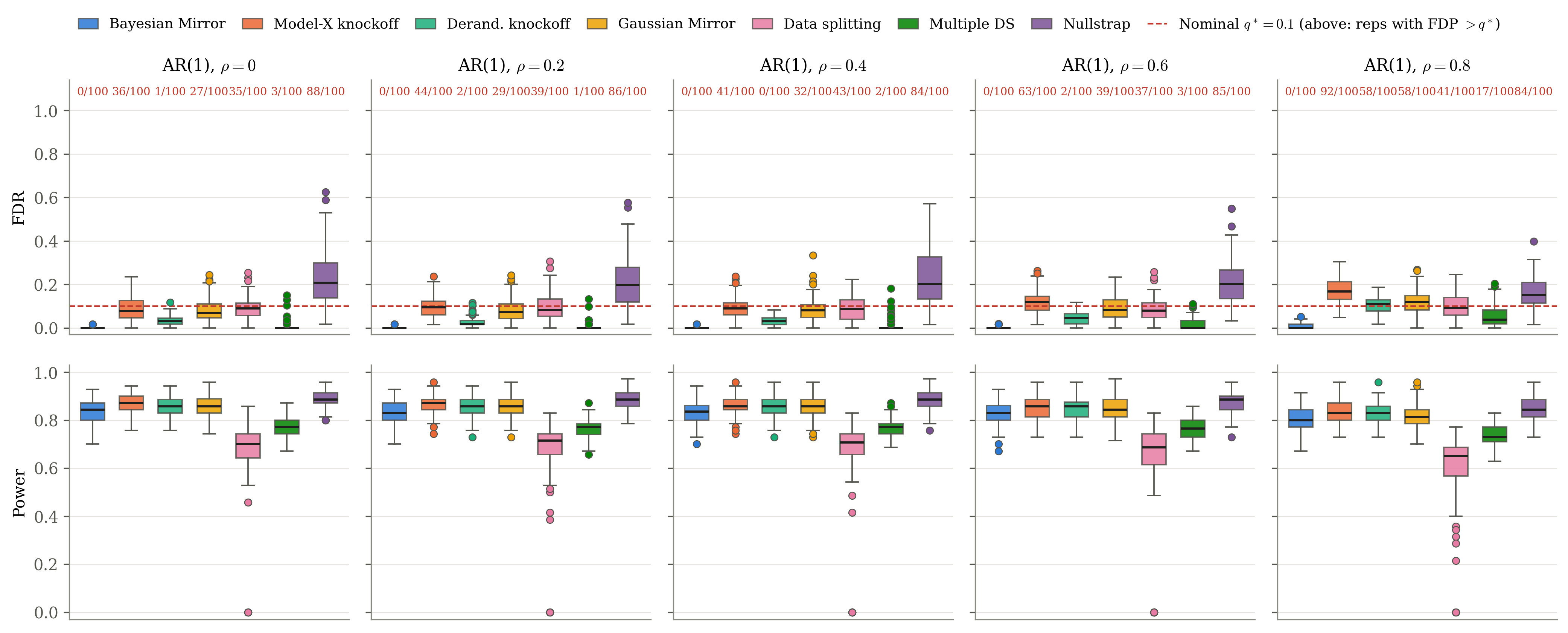}
    \caption{Comparison of FDR and power under autoregressive correlation AR(1).}
    \label{fig:AR}
\end{figure}

Figure~\ref{fig:AR} reports the realised FDP and power across the $100$ replications. The Bayesian mirror keeps its FDP below the target $q^\star=0.1$ in every replication of every setting, with an empirical FDR at most $0.008$. Nullstrap exceeds $q^\star$ throughout, with an FDR between $0.17$ and $0.22$, and since this inflation is already present at $\rho=0$, it is driven by the dimension of the problem rather than by the dependence among the predictors. For $\rho\le0.4$, all other methods control the FDR, and the Bayesian mirror attains power between $0.83$ and $0.84$, within $0.03$ of the two knockoff procedures and the Gaussian mirror, and well above multiple and single data splitting.

The advantage of the Bayesian mirror emerges under stronger dependence. At $\rho=0.6$, model-X knockoffs exceed the nominal level with an FDR of $0.12$. Among the methods that control the FDR, the Bayesian mirror attains power $0.83$, within $0.02$ of derandomised knockoffs and the Gaussian mirror, and well above multiple and single data splitting with power of $0.76$ and $0.60$ respectively. At $\rho=0.8$, derandomised knockoffs, Gaussian mirror and model-X knockoffs exceed the nominal level, with corresponding FDRs of $0.11$, $0.12$ and $0.17$ and exceedances in $58$, $58$ and $92$ replications, respectively. The Bayesian mirror is the most powerful among the methods that control the FDR, with power $0.80$ against $0.73$ of multiple data splitting and $0.60$ of single data splitting, at an FDR of $0.008$. Its power declines by only $0.035$ from $\rho=0$ to $\rho=0.8$.

\paragraph{Block correlation.}

\begin{figure}
    \centering
    \includegraphics[width=1\linewidth]{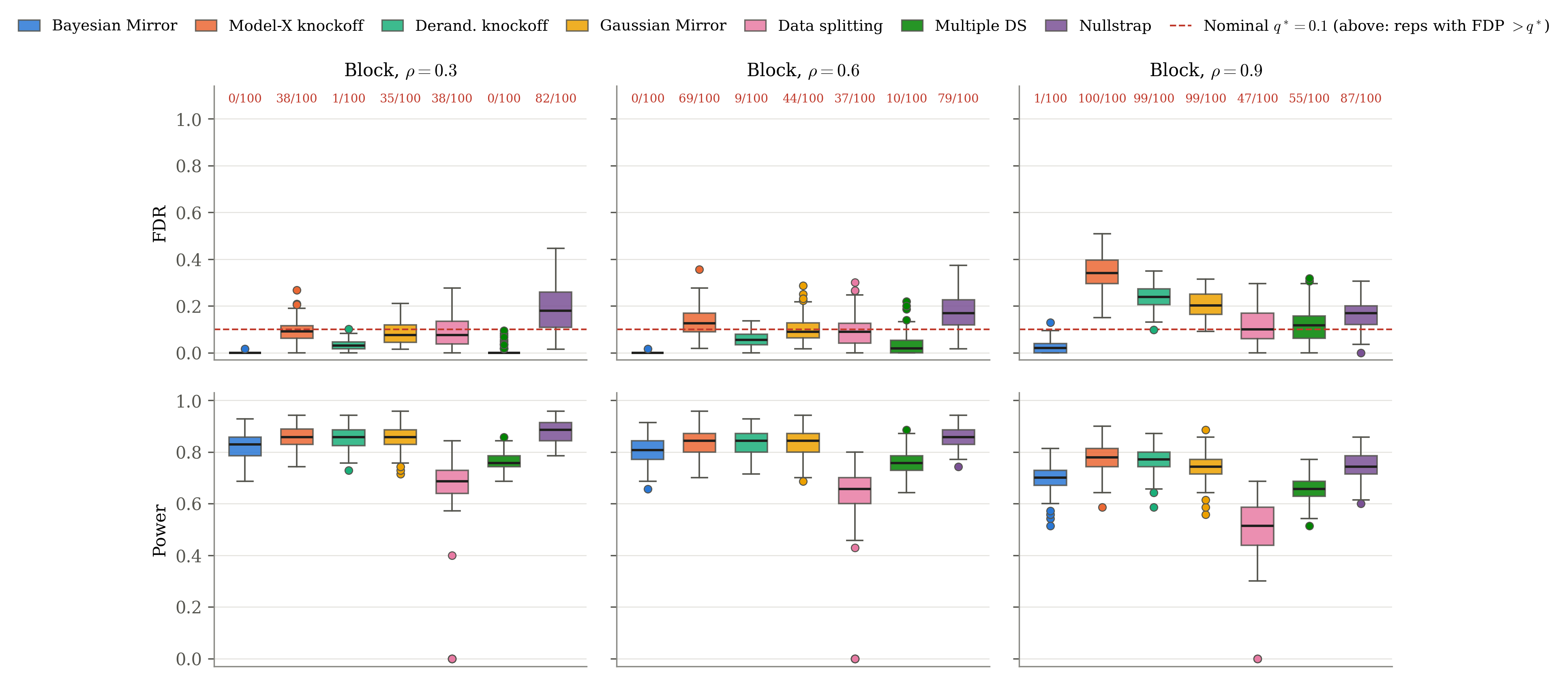}
    \caption{Comparison of FDR and power under block correlation.}
    \label{fig:block}
\end{figure}

Figure~\ref{fig:block} reports the results for the block correlation setting. The Bayesian mirror attains an empirical FDR of at most $0.024$, and its FDP exceeds $q^\star$ in only one single replication across all three levels of correlation $\rho$. Nullstrap exceeds the nominal level throughout, with an FDR between $0.16$ and $0.19$, and the FDR of model-X knockoffs, derandomised knockoffs and the Gaussian mirror increases with $\rho$. At $\rho=0.3$, all other methods control the FDR on average, and the Bayesian mirror attains power $0.82$, against $0.85$ of the two knockoff procedures and the Gaussian mirror, $0.76$ of multiple data splitting and $0.60$ of single data splitting. At $\rho=0.6$, model-X knockoffs and the Gaussian mirror exceed the nominal level with an FDR of $0.13$ and $0.105$ respectively. Among the methods that control the FDR, the Bayesian mirror is second in power with $0.81$, behind derandomised knockoffs with $0.83$, and ahead of multiple and single data splitting with $0.76$ and $0.58$, respectively.

At $\rho=0.9$, every null predictor in the block of a signal is strongly correlated with it, and the Bayesian mirror is the only method that controls the FDR, at $0.024$ with power $0.70$. The FDR is $0.34$ for model-X knockoffs, $0.24$ for derandomised knockoffs and $0.21$ for the Gaussian mirror, whose FDP exceeds $q^\star$ in all replications. Nullstrap attains an FDR of $0.16$, and multiple and single data splitting exceed the nominal level with $0.12$ and $0.11$, respectively. The methods with higher power 
sacrifice the FDR control 
between $0.16$ and $0.34$.

\paragraph{Proxy twins.}

\begin{figure}
    \centering
    \includegraphics[width=1\linewidth]{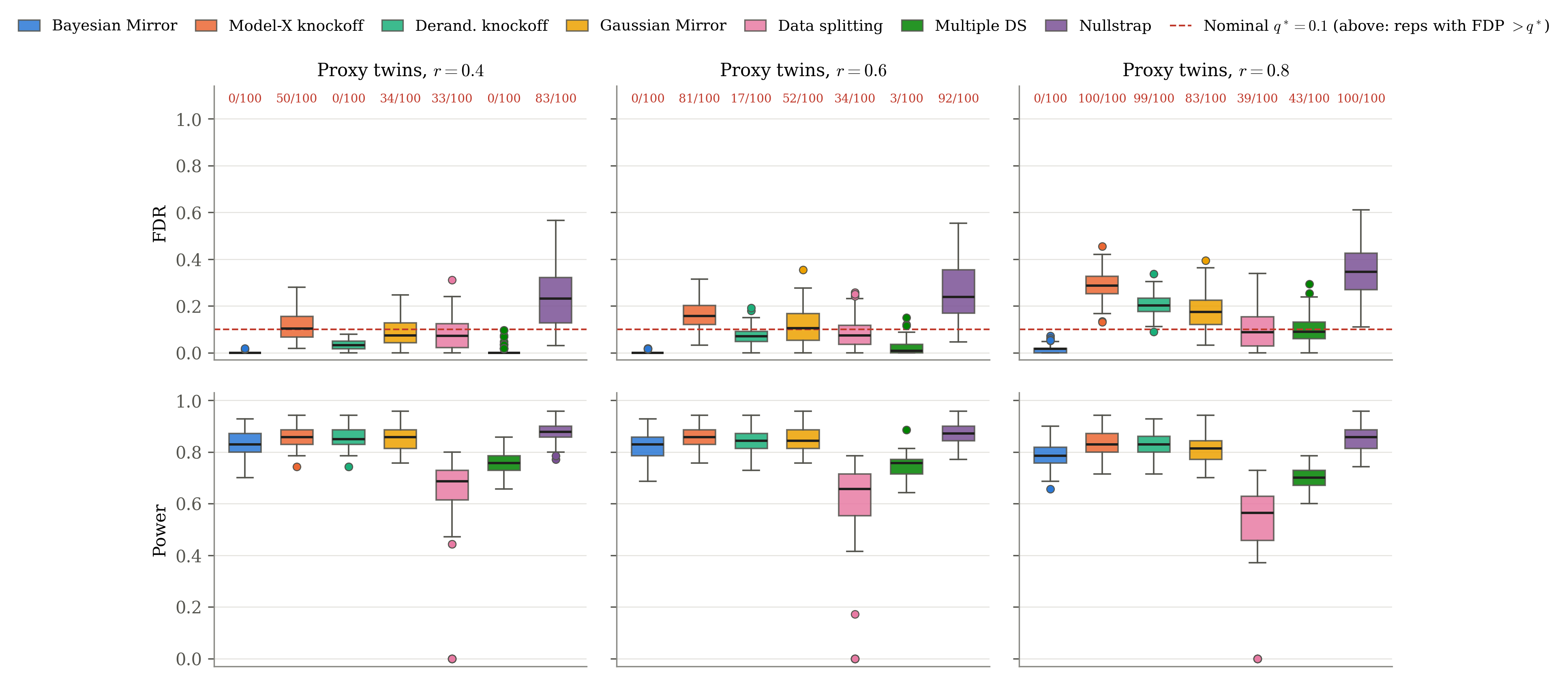}
    \caption{Comparison of FDR and power under the proxy twins setting.}
    \label{fig:twins}
\end{figure}

Figure~\ref{fig:twins} reports the results for the proxy twins setting. The Bayesian mirror keeps its FDP below $q^\star$ in every replication of all three different $r$ values, with an empirical FDR of at most $0.015$. 
Nullstrap exceeds the nominal level throughout, with an FDR between $0.24$ and $0.35$, and the FDR of model-X knockoffs, derandomised knockoffs and the Gaussian mirror increases with $r$. At $r=0.4$, the methods that control the FDR on average are derandomised knockoffs and the Gaussian mirror, with power $0.85$, the Bayesian mirror with $0.83$, and multiple and single data splitting with $0.76$ and $0.59$, respectively. At $r=0.6$, model-X knockoffs and the Gaussian mirror exceed the nominal level, with FDRs of $0.16$ and $0.12$ respectively. Among the methods that control the FDR, the Bayesian mirror is second in power with $0.82$, behind derandomised knockoffs with $0.85$, whose FDP however exceeds $q^\star$ in 17 out of 100 replications.

At $r=0.8$, every twin is a strong proxy for its signal, and model-X knockoffs, derandomised knockoffs and the Gaussian mirror lose FDR control, with FDR of $0.29$, $0.21$ and $0.18$ and exceedances in 100, 99 and 83 replications, respectively. Multiple data splitting attains an FDR of $0.10$, at the nominal level, with 43 exceedances. The Bayesian mirror is the most powerful one among the methods that control the FDR, with power $0.79$ against $0.70$ of multiple data splitting and $0.49$ of single data splitting, at an FDR of $0.015$. The methods with higher power 
sacrifice the FDR control between $0.18$ and $0.35$.

\subsection{Semi-synthetic Application on RNA Expression Data}

We apply the Bayesian mirror and competitive methods to the TCGA glioblastoma multiforme (GBM) gene expression data \citep{TCGA2008,Brennan2013gbm}, profiled on the Affymetrix HT\_HG-U133A platform at the Broad Institute. The data comprise $\log_2(\cdot)$ expression values for $12042$ genes across $538$ samples ($528$ primary tumours and $10$ solid-tissue normals). The data are publicly available from the Broad Institute GDAC Firehose (\url{https://gdac.broadinstitute.org}). We use all samples and genes, standardising each gene, so that the design matrix $X$ has $n=538$ rows and $p=12042$ columns. The design is held fixed across replications, so that the correlation structure is that of the real expression data, while the response is generated synthetically. In each of the $100$ replications, we draw $p_1=30$ genes uniformly at random, set their coefficients $\beta_j^0\sim N(0,1)$ and $\beta_j^0=0$ for all other genes, and generate $y=X\beta^0+\epsilon,\; \epsilon\sim N_n(0,I_n)$. All methods target the FDR at $q^\star=0.1$. For the multi-split Bayesian mirror, we use $L=20$ partitions, a screening fraction of $n_1/n=0.8$, a per-iterate level of $q=0.07$ and $T=1000$ Monte Carlo iterates. We use $q=0.07$ rather than the $q=0.03$ of the simulations, since an iterate can select only if $R\ge1/q$, which at $q=0.03$ exceeds the $p_1=30$ signals, whereas $q=0.07$ requires $15$ selections and still satisfies \eqref{eq:alpha-condition}.

Table~\ref{tab:rnaseq} reports the results for the TCGA data. The Gaussian mirror, data splitting, multiple data splitting, and nullstrap fail to control the FDR, 
with empirical FDRs $0.202$, $0.128$, $0.169$, and $0.114$ and their FDP exceeds $q^\star$ in $73$, $51$, $68$, and $45$ out of the $100$ replications, respectively. Of the three methods whose FDR remains below the nominal level, derandomised knockoffs make no selections at all and model-X knockoffs attain power $0.238$. The Bayesian mirror attains power $0.441$, which almost doubles that of model-X knockoffs and above that of single data-splitting, with an empirical FDR of $0.001$ and no replication in which the FDP exceeds $q^\star$. 

Under the strong and irregular correlation of real expression data, the Bayesian mirror is the only method that combines strict control of the false discovery proportion with substantial power. This is consistent with the simulation studies, in which the FDR of the competing methods increased with the level of the dependence, while that of the Bayesian mirror remained close to zero in every setting. In the most strongly correlated simulation settings and in the expression data, most competing methods lose FDR control, whereas the Bayesian mirror retains it with the highest power among the methods that control the FDR. This robustness is consistent with its construction, since the predictors are screened on one part of the data and each mirror statistic is computed on independent observations, adjusted for all screened predictors, so that a null predictor correlated with a signal cannot be confirmed by the data that selected it.

\begin{table}[t]
\centering
\caption{Semi-synthetic study on the TCGA GBM expression data ($n = 538$, $p = 12042$, $p_1 = 30$, $q^\star = 0.1$, $100$ replications). Standard errors in parentheses; the last column counts replications with FDP $> q^\star$ (i.e., exceedance).}
\label{tab:rnaseq}
\begin{tabular}{lccc}
\toprule
Method & FDR ($q^\star = 0.1$) & Power & FDP $> q^\star$ \\
\midrule
Model-X knockoffs        & 0.018 (0.006) & 0.238 (0.024) & 7/100 \\
Derandomised knockoffs   & 0.000 (0.000) & 0.000 (0.000) & 0/100 \\
Gaussian mirror          & 0.202 (0.012) & 0.634 (0.007) & 73/100 \\
Data splitting           & 0.128 (0.012) & 0.409 (0.016) & 51/100 \\
Multiple data splitting  & 0.169 (0.013) & 0.568 (0.007) & 68/100 \\
Nullstrap                & 0.114 (0.011) & 0.631 (0.008) & 45/100 \\
{\bf Bayesian mirror}          & 0.001 (0.000) & 0.441 (0.026) & 0/100 \\
\bottomrule
\end{tabular}
\end{table}

\subsection{Computational cost}\label{sec:time}

We timed each method on a single replication of the autoregressive setting with $\rho=0.4$, $n=800$, $p=6000$ and $p_1=70$, with all tuning parameters the same as in Section~\ref{sec:sim}, except that multiple data splitting uses $20$ splits, matching the $L=20$ partitions of the Bayesian mirror for fairness. 
Table~\ref{tab:time} reports the run times together with the selections on this replication.

\begin{table}[htbp]
\centering
\caption{Run time in seconds and selections on one replication of the autoregressive setting with $\rho=0.4$, $n=800$, $p=6000$ and $p_1=70$.} 
\label{tab:time}
\begin{tabular}{lrcrr}
\toprule
Method & Time (s) & Selection & FDP & Power \\
\midrule
\textbf{Bayesian mirror}          & \textbf{51.3}   & 60 & 0.000 & 0.857 \\
Data splitting           & 65.4   & 59 & 0.075 & 0.757 \\
Nullstrap                & 108.3   & 69 & 0.116 & 0.871 \\
Model-X knockoffs        & 182.1  & 62 & 0.032 & 0.857 \\
Derandomised knockoffs   & 247.2  & 61 & 0.016 & 0.857 \\
Multiple data splitting  & 1313.5 & 56 & 0.000 & 0.800 \\
Gaussian mirror          & 3497.4 & 67 & 0.075 & 0.886 \\
\bottomrule
\end{tabular}
\end{table}

The Bayesian mirror is the fastest method, although it combines $L=20$ partitions with $T=1000$ posterior iterates, and it is even faster than single data splitting, which uses one partition. 
Almost all of its cost lies in the $20$ cross-validated lasso screenings, while the inference stage over all partitions takes a fraction of a second. This follows from Lemma~\ref{lem:2} and Proposition~\ref{prop:suff}. By Lemma~\ref{lem:2}, each perturbation combines a uniform direction in the residual space $\mathrm{col}(H)$, shared by all screened predictors, with a component $\gamma_j$ along the predictor-specific direction $\tilde q_j$. By Proposition~\ref{prop:suff}, $\gamma_j$ and $w_j^\top y$ are drawn in closed form from $\hat\beta_j$, $\Omega_{jj}$ and $\|Hy\|^2$, which follow from a single factorisation of $X_{\hat J}^{(2)\top} X_{\hat J}^{(2)}$, without constructing a basis of $\mathrm{col}(H)$ or the perturbations $w_j$. Since the conditional posterior is available in closed form, the $T$ iterates are drawn independently in a single vectorised step, without Markov chain Monte Carlo, burn-in or convergence diagnostics. In contrast, the Gaussian mirror refits the lasso for each of the $p$ predictors, and multiple data splitting repeats a ten-fold cross-validated lasso $20$ times. The cost of the Bayesian mirror is therefore that of $L$ lasso fits, and it depends on $p$ only through the screening step.

\subsection{Sensitivity to the screening fraction and per-iterate level}\label{sec:abl}

The performance of the Bayesian mirror, as implemented in Algorithm~\ref{alg:bm}, depends on two tuning parameters. The screening fraction governs the trade-off between the quality of the screened set and the sample size $n_2$ left for inference, and the per-iterate level $q$ sets the threshold applied at each Monte Carlo iterate before aggregation. We assess both on the AR(1) setting with $\rho=0.4$, $n=800$, $p=6000$ and $p_1=70$, over $100$ replications. We vary the screening fraction over $\{0.4,0.5,\dots,0.8\}$ and the per-iterate level over $q\in\{0.02,0.03,\dots,0.08\}$. All other settings are as in the simulation setting in Section~\ref{sec:sim}. Within each replication, the same data and partitions are used for every value of $q$, so that differences across $q$ reflect the per-iterate threshold alone. Table~\ref{tab:ablation} reports the power for each combination.

The empirical FDR does not exceed $0.001$ in any of the $35$ configurations, so neither parameter affects error control in practice, and both act on power alone. Power increases with the screening fraction at every level, from $0.617$ to $0.833$ at $q=0.03$. A larger screening subset retains more of the active set, from $41.5$ to $59.7$ of the $70$ signals on average, and this gain outweighs the smaller sample left for inference. The gain diminishes beyond a screening fraction of $0.6$, once the screen retains most of the active set. For screening fractions of at least $0.6$, power is highest at $q=0.02$ or $q=0.03$, decreases gradually up to $q=0.07$ and sharply at $q=0.08$. At smaller screening fractions power also collapses at $q=0.02$. Since $\widehat{\mathrm{FDP}}(s)\le q$ requires $R(s)\ge1/q$, an iterate at $q=0.02$ selects only if at least $50$ variables exceed the threshold, whereas a screening fraction of $0.4$ retains $41.5$ signals on average. A larger per-iterate level admits more null variables at each iterate, which enlarges the denominator of the per-iterate e-values and dilutes the evidence accumulated by the signals. The configuration used in the simulations in Section~\ref{sec:sim}, a screening fraction of $0.8$ with $q=0.03$, attains the highest power.

\begin{table}[t]
\centering
\caption{Sensitivity of the Bayesian mirror's power to the screening fraction and the per-iterate level $q$ on the AR(1) setting with $\rho = 0.4$ ($n = 800$, $p = 6000$, $p_1 = 70$, $q^\star = 0.1$, $100$ replications).}
\label{tab:ablation}
\begin{tabular}{lccccccc}
\toprule
 & \multicolumn{7}{c}{Per-iterate level $q$} \\
\cmidrule(lr){2-8}
Screening fraction & 0.02 & \textbf{0.03} & 0.04 & 0.05 & 0.06 & 0.07 & 0.08 \\
\midrule
0.4 & 0.032  & 0.617 & 0.613 & 0.572  & 0.538 & 0.455  & 0.056  \\
0.5 & 0.706 & 0.747  & 0.727 & 0.710  & 0.682  & 0.647  & 0.276  \\
0.6 & 0.797  & 0.799  & 0.782  & 0.773  & 0.755  & 0.735  & 0.596  \\
0.7 & 0.819  & 0.820  & 0.807  & 0.800  & 0.786  & 0.765  & 0.673  \\
\textbf{0.8} & 0.832  & \textbf{0.833}  & 0.822  & 0.815  & 0.803  & 0.787  & 0.696  \\
\bottomrule
\end{tabular}%
\end{table}

\section{Conclusion}
\label{sec:conc}
We have developed the Bayesian mirror, a geometric construction of mirror variables that uses a cut posterior distribution to generate the mirror statistics. The perturbations admissible in the mirror construction form a sphere of fixed radius in the residual space of the remaining predictors. Drawing them uniformly under a cut posterior that keeps them independent of the response makes the mirror statistic exactly symmetric under the null at every Monte Carlo iterate. Aggregating the iterates through the e-BH procedure yields a single selection rule with global FDR control under a spectral condition governed by the partial rather than the marginal correlations of the design, together with power guarantees at the per-iterate and aggregated levels.

The method has two practical strengths. The first is speed. Each posterior draw depends on the data only through a single least-squares fit, so inference adds negligible cost and, in high dimensions, the run time is essentially that of the screening step. The Bayesian mirror was the fastest of the methods compared, and more than sixty times faster than the Gaussian mirror, which refits the model for each predictor. The second is robustness to strong dependence in high dimensions. In the strongly correlated simulation settings and in the semi-synthetic study on gene expression data, most competing methods failed to control the FDR, whereas the Bayesian mirror kept the empirical FDR well below the target level in every setting. Among the methods that controlled the FDR in these settings, the Bayesian mirror had the highest power.

The stability of the procedure has two sources. The first is the e-BH aggregation. Averaging the e-values over Monte Carlo iterates and over independent partitions removes the dependence of the selection on a particular perturbation draw and on a particular split, while preserving the validity of the e-values. The second is the separation of screening and inference. Since the two stages use disjoint subsets of the data, the screening event leaves the distribution of the noise in the inference stage unchanged, so the exogeneity of the perturbation and the null symmetry are preserved without any correction for selection. This separation is a plausible reason for its FDR control under strong dependence. In the Gaussian mirror and in data splitting, the mirror statistic also depends on the data used for selection, so strong correlation between null and active predictors can carry the selection event into the null statistics. However, the same two sources account for the power gap relative to single-run methods. Since each e-value is bounded, e-BH can make a selection only when the number of strong signals is large relative to the inverse target level, so at very strict levels or with few signals, no variable can be selected regardless of signal strength, a limitation shared by derandomised knockoffs. Averaging also spreads the contribution of each null across iterates and partitions, so the realised FDR falls well below the target and part of the error budget goes unused. Moreover, each stage uses only part of the sample, so the mirror coefficients are estimated from fewer observations than in methods that screen and infer on the full data. Boosting the e-values before aggregation \citep{Wang2022False-Discovery,Lee2024} may recover part of the unused budget without compromising validity.

In high dimensions, the FDR guarantee further rests on sure screening. This requirement appears stronger than necessary, since in the semi-synthetic study each screen retained only part of the active set, yet the realised FDP never exceeded the target. This suggests that a guarantee tolerating a limited number of omitted signals is attainable. Removing screening altogether would avoid both the reduced sample size and the reliance on sure screening, but the admissible sphere is empty once the number of predictors exceeds the sample size. A shrinkage prior on the nuisance coefficients, combined with perturbations drawn in the residual space of a low-dimensional approximation of the remaining predictors, is a possible route, although the null symmetry would have to be re-established. 

Finally, the construction is specific to the Gaussian linear model, in which conjugacy yields the closed-form two-dimensional update and the orthogonality of Lemma~\ref{th:1} makes the posterior of the mirror pair diagonal. For generalised linear models, data augmentation offers a natural extension. Under P\'olya--Gamma augmentation for logistic regression \citep{Polson2013}, for example, the likelihood is conditionally Gaussian with observation weights, and the mirror construction could be carried out in the weighted residual space. In this case, the weights are learned from the response, so the exogeneity of the perturbation, and hence the null symmetry, must be re-established.

\newpage
\bigskip
\begin{center}
{\large\bf SUPPLEMENTARY MATERIAL}
\end{center}






\subsection*{Auxiliary Lemmas}

\begin{lemma}\label{lem:3}
    For all $u,v\in\mathbb{R}$, 
    \begin{equation*}
        |u+v|-|u-v|=2\,\mathrm{sign}(uv)\min(|u|,|v|),
    \end{equation*}
    and consequently, for every $s>0$,
    \begin{equation*}
        \big\{|u+v|-|u-v|>s\big\}=\Big\{u>\frac{s}{2},\,v>\frac{s}{2}\Big\}\cup \Big\{u<-\frac{s}{2},\,v<-\frac{s}{2} \Big\}.
    \end{equation*}
\end{lemma}

\begin{proof}
    If $u,v\ge0$ then $|u+v|-|u-v|=u+v-|u-v|=2\min(u,v)$, and $\mathrm{sign}(uv)=1$ unless $uv=0$, in which case both sides vanish. The case $u,v\le0$ follows by the evenness of both sides in $(u,v)\mapsto(-u,-v)$. If $u\ge0\ge v$, write $v=-\nu$, with $\nu\ge0$. Then $|u+v|-|u-v|=|u-\nu|-(u+\nu)=-2\min(u,\nu)=-2\min(|u|,|v|)$, and $\mathrm{sign}(uv)=-1$ unless $uv=0$, in which case both sides vanish. The case $u\le0\le v$ is symmetric. For the second claim, a value exceeding $s>0$ forces $\mathrm{sign}(uv)=1$ and $\min(|u|,|v|)>s/2$, that is, $u$ and $v$ of the same sign with both magnitudes exceeding $s/2$. Conversely, any such pair gives the value $2\min(|u|,|v|)>s$.
\end{proof}

\subsection*{Proofs}\label{ap:proofs}

\begin{proof}[Proof of Lemma~\ref{th:1}]
Projecting out $X_{-j}$, the relevant inner product given $X_{-j}$ is the partial covariance,
\(
(X_j^+)^\top H_{-j} X_j^- = (X_j + w_j)^\top H_{-j} (X_j - w_j),\)
which expands to
\(X_j^\top H_{-j} X_j - X_j^\top H_{-j} w_j + w_j^\top H_{-j} X_j - w_j^\top H_{-j} w_j.\)
Since $H_{-j}$ is symmetric, the middle terms cancel, and $w_j^\top H_{-j}w_j=\|H_{-j}w_j\|^2$ because $H_{-j}$ is idempotent. Since $\alpha_j=X_j^\top H_{-j}X_j$, the partial covariance reduces to 
\[(X_j^+)^\top H_{-j} X_j^- = \alpha_j - \|H_{-j}w_j\|^2\]
and the stated equivalence is immediate. 
\end{proof}

\begin{proof}[Proof of Lemma~\ref{lem:1}]
    Write $\iota_j$ for the $j$-th canonical basis vector. Since $H_{-j}X_j=X_j-(I_n-H_{-j})X_j$ lies in $\mathrm{col}(X)$ and is orthogonal to $\mathrm{col}(X_{-j})$, it spans the one-dimensional space $\mathrm{col}(X)\cap\mathrm{col}(X_{-j})^\perp$, and it is the non-zero element $v$ of that space with $X_j^\top v=\|v\|^2$. The vector $X\Omega_{\cdot j}$ satisfies $X^\top(X\Omega_{\cdot j})=\iota_j$, so it lies in the same space, with $X_j^\top X\Omega_{\cdot j}=1$ and 
    \[\|X\Omega_{\cdot j}\|^2=\Omega_{\cdot j}^\top X^\top X\,\Omega_{\cdot j}=(\Omega X^\top X\Omega)_{jj}=\Omega_{jj}.\]
    Hence $v=X\Omega_{\cdot j}/\Omega_{jj}$ satisfies $X_j^\top v=\|v\|^2=1/\Omega_{jj}$, which gives $H_{-j}X_j=X\Omega_{\cdot j}/\Omega_{jj}$ and
    $\alpha_j=\|H_{-j}X_j\|^2=1/\Omega_{jj}$. The Gram identity follows since $\tilde q_j^\top\tilde q_k=\Omega_{\cdot j}^\top X^\top X \Omega_{\cdot k}/\sqrt{\Omega_{jj}\Omega_{kk}}=\Omega_{jk}/\sqrt{\Omega_{jj}\Omega_{kk}}$.
\end{proof}

\begin{proof}[Proof of Lemma~\ref{lem:2}]
    Both $H_{-j}$ and $H+\tilde q_j\tilde q_j^\top$ are symmetric idempotents, the latter because $\tilde q_j$ is a unit vector orthogonal to $\mathrm{col}(H)$, and they have the same range. Indeed, $\mathrm{col}(H_{-j})$ is the orthogonal complement of $\mathrm{col}(X_{-j})$, which contains $\mathrm{col}(H)=\mathrm{col}(X)^\perp$ together with the single additional direction $\tilde q_j\in\mathrm{col}(X)\cap\mathrm{col}(X_{-j})^\perp$. Orthogonality of $\tilde q_j$ to $\mathrm{col}(H)$ is immediate from $\tilde q_j\in\mathrm{col}(X)$, and comparing dimensions gives $\dim\mathrm{col}(H)=r-1=n-p$. Fix $j$. The decomposition (\ref{eq:w_decomp}) is the resolution of the unit vector $\tilde w_j\in\mathrm{col}(H_{-j})$ into its component along $\tilde q_j$, of length $\gamma_j$, and its component along $\mathrm{col}(H)$, of length $\sqrt{1-\gamma_j^2}$, normalised to the unit vector $\tilde u_j$. For a uniform point on a sphere, the normalised projection onto a subspace is uniform on the unit sphere of that subspace and is independent of the component orthogonal to it. Applied to the orthogonal splitting above, this gives $\tilde u_j\sim\mathrm{Unif}\{u\in\mathrm{col}(H):\|u\|=1\}$ independent of $\gamma_j$. For mutual independence across $j$, conditionally on $X$ the subspace $\mathrm{col}(H)$ and each direction $\tilde q_j$ are fixed, so the pair $(\tilde u_j,\gamma_j)$ is a measurable function of $v_j$ alone. Since the $v_j$ are drawn independently, the $p$ pairs are mutually independent.
\end{proof}

\begin{proof}[Proof of Proposition~\ref{prop:suff}]
    By Lemma~\ref{lem:1}, $\tilde q_j^\top y=\Omega_{\cdot j}^\top X^\top y/\sqrt{\Omega_{jj}}=\hat\beta_j/\sqrt{\Omega_{jj}}$ and $\sqrt{\alpha_j}=\Omega_{jj}^{-1/2}$, which gives the first identity, and the second follows from \eqref{eq:hy_decomp}. Since $U_H$ is an orthonormal basis of $\mathrm{col}(H)$, $\|U_H^\top y\|=\|Hy\|$. As $\xi_j\sim N(0,I_{n-p})$ is rotation invariant, its coordinate $h_j$ along the unit vector $U_H^\top y/\|Hy\|$ is standard normal and independent of the squared norm $\|\xi_j\|^2-h_j^2\sim\chi^2_{n-p-1}$ of its orthogonal component. Substituting $\xi_j^\top U_H^\top y=\|Hy\|\,h_j$ and the first identity into $w_j^\top y$ under \eqref{eq:w_sampler} gives \eqref{eq:wy}. Finally, \eqref{eq:c_def}, \eqref{eq:d_kappa}, \eqref{eq:sigma_reduced} and \eqref{eq:lem_rand} depend on $w_j$ only through $\gamma_j$ and $w_j^\top y$, and on the data only through $\alpha_j=1/\Omega_{jj}$, $\sqrt{\alpha_j}\,\tilde q_j^\top y$ and $\|H_{-j}y\|^2$. By the first two identities, these are functions of $\hat\beta_j$, $\Omega_{jj}$ and $\|Hy\|^2$, and each draw requires a fixed number of scalar operations.
\end{proof}

\begin{proof}[Proof of Lemma~\ref{lemma1}]
   Fix a realisation $w_j^{(t)}=w_j$ and an iterate $t$. Since both are held fixed we drop the superscript $(t)$. We write $\mathbb{P}_{w_j}$ for probability conditional on $(X,w_j)$, under which every vector built from $w_j$ is non-random. By Remark~\ref{rem:nondegeneracy} we may assume $c_j^{\pm}\neq0$, so that $a_j^{\pm}>0$.

   The conditional posteriors \eqref{eq:beta_reduced}--\eqref{eq:sigma_reduced} apply, with $\beta_j^+\perp\!\!\!\perp\beta_j^-$ given $(\sigma_j^2,w_j,y)$ by Lemma~\ref{th:1} and with inverse-gamma rate $b_n$ of \eqref{eq:sigma_reduced}.
   Since $j\in\mathcal{H}_0$ and $H_{-j}X_{-j}=0$ we have $H_{-j}y=H_{-j}\epsilon$, so by \eqref{eq:d_kappa} and the symmetry and idempotence of $H_{-j}$,
   \begin{equation}\label{eq:lem_d}
       d_j^\pm=(c_j^\pm)^\top H_{-j}\epsilon=(c_j^{\pm})^\top\epsilon.
   \end{equation}

   Under the cut posterior, $v_j$ is sampled first from $\mathrm{Unif}(\mathbb{S}_j)$, independently of $y$, hence $w_j$ is independent of $\epsilon$. Therefore, conditioning on $w_j$ does not alter the distribution of $\epsilon$,
   \begin{equation*}
       \epsilon\mid w_j\sim N(0,\sigma_0^2I_n).
   \end{equation*}
   The vectors $c_j^+/\|c_j^+\|$ and $c_j^-/\|c_j^-\|$ are orthonormal in $\mathrm{col}(H_{-j})$ by Lemma~\ref{th:1}. We complete them to an orthonormal basis and let $\|\tilde{\epsilon}_j\|^2$ denote the squared norm of the component of $H_{-j}\epsilon$ along the remaining $r-2$ directions. Using $\|c_j^\pm\|^2=2a_j^\pm$,
   \begin{equation}\label{eq:lem_decomp}
   \|H_{-j}y\|^2=\frac{(d_j^+)^2}{2a_j^+}+\frac{(d_j^{-})^2}{2a_j^-}+\|\tilde{\epsilon}_j\|^2. 
   \end{equation}
   Since the coordinates of $H_{-j}\epsilon$ in this basis are i.i.d.\ $N(0,\sigma_0^2)$, $d_j^+$, $d_j^-$ and $\|\tilde{\epsilon}_j\|^2$ are mutually independent, with $d_j^{\pm}\sim N(0,2\sigma_0^2a_j^\pm)$.

   The draw is given by \eqref{eq:lem_rand}. Substituting \eqref{eq:lem_decomp} into the rate $b_n$ of \eqref{eq:sigma_reduced} and using $1/2a_j^\pm-1/\kappa_j^\pm=\tau^{-2}/2a_j^\pm\kappa_j^\pm$,
   \begin{equation}\label{eq:lem_even}
       b_n=b_0+\frac{\tau^{-2}}{2}\left[\frac{(d_j^+)^2}{2a_j^+\kappa_j^+}+\frac{(d_j^-)^2}{2a_j^-\kappa_j^-}\right]+\frac{\|\tilde{\epsilon}_j\|^2}{2},
   \end{equation}
   so $\widetilde M_j$ is a measurable function of $\left(d_j^+,d_j^-,\|\tilde{\epsilon}_j\|^2,z_j^+,z_j^-,G_j\right)$.

   Define the involution
   \[T:\left(d_j^+,d_j^-,\|\tilde{\epsilon}_j\|^2,z_j^+,z_j^-,G_j\right)\longmapsto\left(d_j^+,-d_j^-,\|\tilde{\epsilon}_j\|^2,z_j^+,-z_j^-,G_j\right).\]
   These six quantities are mutually independent under $\mathbb{P}_w$, and the laws of $d_j^-$ and $z_j^-$ are centred Gaussian, hence symmetric about zero. The remaining four are unchanged. Therefore $T$ preserves $\mathbb{P}_w$. By (\ref{eq:lem_even}), $b_n$ depends on $d_j^-$ only through $(d_j^-)^2$, so $b_n\circ T=b_n$ and, since $G_j\circ T=G_j$, 
   \begin{equation}\label{eq:lem_scale_inv}
    \sigma_j^2\circ T=\sigma_j^2, \qquad \hat s_j\circ T=\hat s_j,
   \end{equation}
   the latter because $\kappa_j^\pm$ are non-random given $(X,w_j)$. Consequently, $\beta_j^+\circ T=\beta_j^+$, being a function of $d_j^+$, $\sigma_j^2$ and $z_j^+$ alone, while by \eqref{eq:lem_rand}
   \[\beta_j^-\circ T=\frac{-d_j^-}{\kappa_j^-}+\sqrt{\frac{\sigma_j^2}{\kappa_j^-}}\left(-z_j^-\right)=-\beta_j^-.\]
   
   By Lemma~\ref{lem:3}, $M_j=2\,\mathrm{sign}(\beta_j^+\beta_j^-)\min(|\beta_j^+|,|\beta_j^-|)$. The involution preserves $|\beta_j^+|$ and $|\beta_j^-|$ and reverses $\mathrm{sign}(\beta_j^+\beta_j^-)$, so $M_j\circ T=-M_j$, which along with \eqref{eq:lem_scale_inv} gives $\widetilde M_j\circ T=-\widetilde M_j$ pointwise. Since $T$ preserves $\mathbb{P}_w$, for every $s>0$
   \[\mathbb{P}_{w_j}\!\left(\widetilde M_j\ge s\right)=\mathbb{P}_{w_j}\!\left(\widetilde M_j\circ T\ge s\right)=\mathbb{P}_{w_j}\!\left(-\widetilde M_j\ge s\right)=\mathbb{P}_{w_j}\!\left(\widetilde M_j\le -s\right).\]

\end{proof}

\begin{proof}[Proof of Corollary~\ref{cor:all_perturbations}]
    Under the cut posterior, the perturbations $\{w_k^{(t)}\}_{k=1}^{p}$ are drawn independently from $\bigotimes_{k=1}^{p}\mathrm{Unif}(\mathbb{S}_k)$. From the proof of Lemma~\ref{lemma1}, $\widetilde M_j^{(t)}$ is a measurable function of $(d_j^\pm,\|\tilde{\epsilon}_j\|^2,z_j^\pm,G_j)$, each of which is built from $w_j^{(t)}$, the noise $\epsilon$, and the auxiliary variables indexed by $j$, none of which involves $w_k^{(t)}$ for $k\neq j$. Consequently, conditioning on the additional perturbations $\{w_k^{(t)}\}_{k\neq j}$ does not alter the conditional law of $\widetilde M_j^{(t)}$, so  
    \[\mathbb{P}\!\left(\widetilde M_j^{(t)}\ge s\mid \mathbf{w}\right)=\mathbb{P}\!\left(\widetilde M_j^{(t)}\ge s\mid w_j^{(t)}\right),\]
    and likewise for $\widetilde M_j^{(t)}\le -s$. The claim then follows from Lemma~\ref{lemma1}.
\end{proof}

\begin{proof}[Proof of Lemma~\ref{lem:4}]
    By construction, $\tilde w_j$ is uniform on the unit sphere of the $r$-dimensional subspace $\mathrm{col}(H_{-j})$, and $\tilde q_j$ is a fixed unit vector of that subspace, so $\gamma_j=\langle\tilde w_j,\tilde q_j\rangle$ is distributed as a single coordinate of a uniform point on $\mathbb{S}^{r-1}$. The spherical cap inequality \citep[Lemma~2.2]{Ball1997-Intro} gives $\mathbb{P}(|\gamma_j|>t)\le 2e^{-rt^2/2}$ for $t\in(0,1)$. Taking $t=\{2(K+1)\log p/r\}^{1/2}$, which lies in $(0,1)$ once $r>2(K+1)\log p$, and applying a union bound over $j=1,\dots,p$ yields the stated bound with probability at least $1-2p\cdot p^{-(K+1)}=1-2p^{-K}$.

    On that event, $n-p\gg\log p$ gives $\max_j|\gamma_j|=o(1)$, so $a_j^\star=\alpha_j(1-|\gamma_j|)\ge\alpha_j\{1-o(1)\}\ge c_0\alpha_{\max}\{1-o(1)\}\ge c_0\alpha_{\max}/2$ for sufficiently large $p$, by Assumption~\ref{as:design}. Finally, $a_j^{\max}/a_j^\star=(1+|\gamma_j|)/(1-|\gamma_j|)=1+o(1)$ uniformly in $j$.
\end{proof}

\begin{proof}[Proof of Lemma~\ref{lem:5}]
    By \eqref{eq:lem_rand} and the surrogate coefficients,
    \[
    \beta_j^\pm=\frac{d_j^\pm}{\kappa_j^\pm}+\frac{\sigma_j}{\sigma_0}\frac{\sigma_0z_j^\pm}{\sqrt{\kappa_j^\pm}}, \qquad \beta_j^\pm(\sigma_0)=\frac{d_j^\pm}{\kappa_j^\pm}+\frac{\sigma_0z_j^\pm}{\sqrt{\kappa_j^\pm}}.
    \]
    Write $\Psi(u,v)=|u+v|-|u-v|$, which is positively homogeneous and satisfies $|\Psi(u,v)-\Psi(u',v')|\le2(|u-u'|+|v-v'|)$. Since $\hat s_j=\sigma_j\theta_j$,
    \[
    \widetilde M_j=\frac{\Psi(\beta_j^+,\beta_j^-)}{\sigma_j\theta_j}=\frac{1}{\sigma_0\theta_j}\Psi\!\left(\frac{\beta_j^+}{\sigma_j/\sigma_0},\frac{\beta_j^-}{\sigma_j/\sigma_0}\right),\qquad \frac{\beta_j^\pm}{\sigma_j/\sigma_0}=\frac{d_j^\pm}{\kappa_j^\pm\sigma_j/\sigma_0}+\frac{\sigma_0z_j^\pm}{\sqrt{\kappa_j^\pm}}.
    \]
    The arguments of $\Psi$ in $\widetilde M_j$ and $\widetilde M_j(\sigma_0)$ differ only in the first summand, the variables $z_j^\pm$ cancelling identically, so
    \begin{equation}\label{eq:ap_coupling}
        \big|\widetilde M_j-\widetilde M_j(\sigma_0)\big|\le\frac{2}{\sigma_0\theta_j}\Big|\frac{\sigma_0}{\sigma_j}-1\Big|\left(\frac{|d_j^+|}{\kappa_j^+}+\frac{|d_j^-|}{\kappa_j^-}\right).
    \end{equation}

    Let $g_j^\pm=(c_j^\pm)^\top\epsilon\sim N(0,2\sigma_0^2a_j^\pm)$, so $d_j^\pm=a_j^\pm\beta_j^0+g_j^\pm$. By \eqref{eq:c_def} and $\kappa_j^\pm=2a_j^\pm+\tau^{-2}$ we have $2a_j^\pm\le\kappa_j^\pm$, and 
    \begin{equation}\label{eq:ap_theta}
        \frac{1}{2\kappa_j^{\min}}\le \theta_j^2\le\frac{1}{\kappa_j^{\min}}, \qquad \kappa_j^{\min}=2a_j^\star+\tau^{-2}.
    \end{equation}
    By (\ref{eq:ap_theta}), $1/(\sigma_0\theta_j)\le\sqrt{2\kappa_j^{\min}}/\sigma_0$. For the signal part $a_j^\pm/\kappa_j^\pm\le1/2$ and $\kappa_j^{\min}=2a_j^\star+\tau^{-2}\le2(1+c_v)a_j^\star$ since $a_j^\star\ge\alpha_j/2\ge\min_k\alpha_k/2$ by Assumption~\ref{as:spectral} and $\tau^{-2}\le c_v\min_k\alpha_k$ by Assumption~\ref{as:design}, so 
    \[
    \frac{2}{\sigma_0\theta_j}\cdot\frac{a_j^\pm|\beta_j^0|}{\kappa_j^\pm}\le\frac{|\beta_j^0|\sqrt{2\kappa_j^{\min}}}{\sigma_0}\le\sqrt{2(1+c_v)}\;\frac{|\beta_j^0|\sqrt{a_j^\star}}{\sigma_0}=C\psi_j.
    \]
    For the noise part, $\kappa_j^\pm\ge2a_j^\pm$ and $\kappa_j^{\min}\le\kappa_j^\pm$ give
    \[\frac{2}{\sigma_0\theta_j}\cdot\frac{|g_j^\pm|}{\kappa_j^\pm}\le\frac{2\sqrt{2\kappa_j^{\min}}}{\sigma_0}\cdot\frac{|g_j^\pm|}{\kappa_j^\pm}\le\frac{2\sqrt{2}}{\sigma_0\sqrt{\kappa_j^\pm}}|g_j^\pm|\le\frac{2\sqrt{2}\,|g_j^\pm|}{\sigma_0\sqrt{2a_j^\pm}}.\]
    Each $g_j^\pm/(\sigma_0\sqrt{2a_j^\pm})$ is standard normal, so a union bound over the $2p$ pairs gives
    \begin{equation}\label{eq:ap_gauss}
        \mathbb{P}_\mathbf{w}\Big(\max_{j,\pm}\big|g_j^\pm\big|/\big(\sigma_0\sqrt{2a_j^\pm}\big)>\sqrt{2(K+1)\log p}\Big)\le 2p^{-K},
    \end{equation}
    and on that event
    \begin{equation}\label{eq:ap_bracket}
        \frac{2}{\sigma_0\theta_j}\left(\frac{|d_j^+|}{\kappa_j^+}+\frac{|d_j^-|}{\kappa_j^-}\right)\le C\big(\psi_{\max}+\sqrt{K\log p}\big)\qquad\text{for every }j.
    \end{equation}

    By \eqref{eq:lem_rand}, $(\sigma_j/\sigma_0)^2=b_n/(\sigma_0^2G_j)$ with $G_j\sim\mathrm{Gamma}(a_0+r/2,1)$, and by \eqref{eq:lem_even},
    \begin{equation}\label{eq:ap_bn}
    \frac{b_n}{\sigma_0^2}=\frac{b_0}{\sigma_0^2}+\frac{\chi_j^2}{2\sigma_0^2}+\frac{\tau^{-2}}{2\sigma_0^2}\left[\frac{(d_j^+)^2}{2a_j^+\kappa_j^+}+\frac{(d_j^-)^2}{2a_j^-\kappa_j^-}\right].
    \end{equation}
    Here $\chi_j^2$ is the squared norm of the component of $H_{-j}\epsilon$ orthogonal to $\mathrm{span}\{c_j^+,c_j^-\}$ within $\mathrm{col}(H_{-j})$. Since $H_{-j}X_j=(c_j^++c_j^-)/2$ lies in that span, the decomposition \eqref{eq:lem_decomp} holds with $\chi_j^2$ unchanged when $\beta_j^0\neq0$, so $\chi_j^2\sim\sigma_0^2\chi^2_{r-2}$ for every $j$, null or not. Lemma~1 of \citet{LaurentMassart2000-Adapt} with $x=K\log p$ and a union bound give
    \begin{equation}\label{eq:ap_chisq}
        \mathbb{P}_\mathbf{w}\left(\max_j\left|\frac{\chi_j^2}{\sigma_0^2(r-2)}-1\right|>2\sqrt{\frac{K\log p}{r-2}}+\frac{2K\log p}{r-2}\right)\le2p^{1-K},
    \end{equation}
    so that $\max_j\big|\chi_j^2/(2\sigma_0^2)-r/2\big|\le C\big(\sqrt{rK\log p}+K\log p+1\big)$ on the complementary event.
    For the third term of (\ref{eq:ap_bn}) we use $(d_j^\pm)^2\le2(a_j^\pm)^2(\beta_j^0)^2+2(g_j^\pm)^2$ and bound the two contributions separately. Since $a_j^\pm/\kappa_j^\pm\le 1/2$ and $\psi_j^2=(\beta_j^0)^2a_j^\star/\sigma_0^2$,
    \[
    \frac{\tau^{-2}}{2\sigma_0^2}\cdot\frac{2(a_j^\pm)^2(\beta_j^0)^2}{2a_j^\pm\kappa_j^\pm}=\frac{\tau^{-2}(\beta_j^0)^2}{2\sigma_0^2}\cdot\frac{a_j^\pm}{\kappa_j^\pm}\le\frac{\tau^{-2}\psi_j^2}{4a_j^\star}.
    \]
    For the noise contribution,
     $\kappa_j^\pm\ge\tau^{-2}$ gives, on the event of \eqref{eq:ap_gauss},
    \[
    \frac{\tau^{-2}}{2\sigma_0^2}\cdot\frac{2(g_j^\pm)^2}{2a_j^\pm\kappa_j^\pm}\le\frac{(g_j^\pm)^2}{2\sigma_0^2a_j^\pm}\le2(K+1)\log p.
    \]
    Combining, the third term of \eqref{eq:ap_bn} is at most $\tau^{-2}\psi_{\max}^2/(2\min_ja_j^\star)+4(K+1)\log p$ uniformly in $j$.
    
    For the gamma variable, the Chernoff bounds $\mathbb{P}(G\lessgtr\lambda k)\le(\lambda e^{1-\lambda})^k$ for $G\sim\mathrm{Gamma}(k,1)$, applied with $k=a_0+r/2\ge r/2$ and $\lambda=1\pm C\sqrt{K\log p/r}$, give
    \begin{equation}\label{eq:ap_gamma}
        \mathbb{P}_\mathbf{w}\left(\max_j\left|\frac{G_j}{a_0+r/2}-1\right|>C\sqrt{\frac{K\log p}{r}}\right)\le2p^{1-K},
    \end{equation}
    valid once $r\ge CK\log p$, which holds under Assumption~\ref{as:regime}.

    Collecting the three displays and dividing \eqref{eq:ap_bn} by $r/2$,
    \[
    \max_j\left|\frac{2b_n}{r\sigma_0^2}-1\right| \ \le\ C\left\{\sqrt{\frac{K\log p}{r}}+\frac{1}{r}\left(1+\frac{b_0}{\sigma_0^2}+K\log p\right)+\frac{\tau^{-2}}{r}\,\frac{\psi_{\max}^2}{\min_j a_j^\star} \right\},
    \]
    while \eqref{eq:ap_gamma} and $|(r/2)/(a_0+r/2)-1|\le2a_0/r$ give
    $\big|2G_j/r-1\big|\le C\{\sqrt{K\log p/r}+a_0/r\}$. Since $(\sigma_j/\sigma_0)^2=(2b_n/(r\sigma_0^2))\big/(2G_j/r)$ and $|x^{-1/2}-1|\le|x-1|$ for $x\ge1/2$,
    \begin{equation}\label{eq:ap_rho}
        \mathbb{P}_\mathbf{w}\Big(\max_j\big|\frac{\sigma_0}{\sigma_j}-1\big|>C\omega_p\Big)\le C'p^{1-K},\quad \omega_p=\sqrt{\frac{K\log p}{r}}+\frac{1}{r}\left(1+a_0+\frac{b_0}{\sigma_0^2}+K\log p\right)+\frac{\tau^{-2}}{r}\,\frac{\psi_{\max}^2}{\min_ja_j^\star}.
    \end{equation}

    Let $\mathcal{G}$ be the intersection of the events in \eqref{eq:ap_gauss}, \eqref{eq:ap_chisq}, \eqref{eq:ap_gamma} and \eqref{eq:ap_rho}. By a union bound $\mathbb{P}_\mathbf{w}(\mathcal{G}^c)\le Cp^{1-K}$. On $\mathcal{G}$, combining the pathwise bound \eqref{eq:ap_coupling} with \eqref{eq:ap_bracket} and \eqref{eq:ap_rho},
    \[
    \max_{j\le p}\big|\widetilde M_j-\widetilde M_j(\sigma_0)\big|\ \le\ C\omega_p\big(\psi_{\max}+\sqrt{K\log p}\big)\ =\ \Delta_p,
    \]
    with $C$ depending only on $c_v$, and $\Delta_p\to0$ by Assumption~\ref{as:regime}. Replacing $K$ by $K+1$ throughout yields $\mathbb{P}_\mathbf{w}(\mathcal{A}_p^c)\le Cp^{-K}$, as stated.

\end{proof}

\begin{proof}[Proof of Proposition~\ref{prop:conc}]
    We first show that, for every $s\in\mathbb{R}$ and $\varepsilon>0$,
    \begin{equation}\label{eq:conc}
        \mathbb{P}_\mathbf{w}\Big(\big|V_{\sigma_0}^+(s)-\mathbb{E}_\mathbf{w} V_{\sigma_0}^+(s)\big|>\varepsilon p_0\Big)\le2e^{-\varepsilon^2p_0/2}+2\exp\!\Big(-\frac{\pi\varepsilon^2p_0}{8\Lambda}\Big).
    \end{equation}
    Set $\tilde\epsilon=\epsilon/\sigma_0\sim N(0,I_n)$ and 
    \[
    B_j^\pm=\frac{\beta_j^\pm(\sigma_0)}{\sigma_0\theta_j}=\mu_j^\pm+\varsigma_j^\pm z_j^\pm, \qquad \mu_j^\pm=\frac{a_j^\pm\beta_j^0+\sigma_0(c_j^\pm)^\top\tilde\epsilon}{\sigma_0\theta_j\kappa_j^\pm}, \qquad \varsigma_j^\pm=\frac{1}{\theta_j\sqrt{\kappa_j^\pm}},
    \]
    so that $\widetilde M_j(\sigma_0)=\Psi(B_j^+,B_j^-)$ by homogeneity. Let $F(\tilde\epsilon)=\mathbb{E}[V_{\sigma_0}^+(s)\mid\tilde\epsilon]$; by the tower property $\mathbb{E}_\mathbf{w}F=\mathbb{E}_\mathbf{w}V_{\sigma_0}^+(s)$, and it suffices to bound $V_{\sigma_0}^+-F$ and $F-\mathbb{E}_\mathbf{w}F$ at level $\varepsilon p_0/2$ each.

    Given $\tilde\epsilon$, the quantities $\mu_j^\pm,\varsigma_j^\pm$ are fixed and $B_j^\pm$ depends only on $z_j^\pm$. As the $z_j^\pm$ are mutually independent and independent of $\epsilon$, the indicators $\{\mathbf{1}(\widetilde M_j(\sigma_0)\ge s)\}_{j\in\mathcal{H}_0}$ are conditionally independent and $[0,1]$-valued, so Hoeffding's inequality gives
    \[
    \mathbb{P}\big(|V_{\sigma_0}^+(s)-F(\tilde\epsilon)|>\varepsilon p_0/2\mid\tilde\epsilon\big)\le2\exp\{-2(\varepsilon p_0/2)^2/p_0\}=2e^{-\varepsilon^2 p_0/2},
    \]
    uniformly in $\tilde\epsilon$ and hence after integration.

    Fix $s>0$. By Lemma~\ref{lem:3}, $\{\widetilde M_j(\sigma_0)>s\}=\{B_j^+>s/2,B_j^->s/2\}\cup\{B_j^+<-s/2,B_j^-<-s/2\}$, a union of two disjoint events. Given $\tilde\epsilon$, the $B_j^\pm$ are independent Gaussians, so 
    \begin{equation}\label{eq:ap_F}
        F(\tilde\epsilon)=\sum_{j\in\mathcal{H}_0}\Big\{\bar\Phi(A_j^+)\bar\Phi(A_j^-)+\Phi(\bar A_j^+)\Phi(\bar A_j^-)\Big\},\quad A_j^\pm=\frac{s/2-\mu_j^\pm}{\varsigma_j^\pm},\ \ \bar A_j^\pm=\frac{-s/2-\mu_j^\pm}{\varsigma_j^\pm}.
    \end{equation}
    For $s\le0$, $\{\widetilde M_j(\sigma_0)\ge s\}$ is the complement of $\{-\widetilde M_j(\sigma_0)>-s\}$, which by Lemma~\ref{lem:3} is a union of two disjoint mixed-sign orthants, so $F$ is again a sum of products of Gaussian distribution functions and the gradient bound below applies verbatim.

    Differentiating \eqref{eq:ap_F} and using $\partial A_j^+/\partial\mu_j^+=\partial\bar A_j^+/\partial\mu_j^+=-1/\varsigma_j^+$,
    \[
    \frac{\partial F}{\partial\mu_j^+}=\frac{1}{\varsigma_j^+}\Big\{\phi(A_j^+)\bar\Phi(A_j^-)-\phi(\bar A_j^+)\Phi(\bar A_j^-)\Big\}.
    \]
    Both terms are non-negative and each is at most  $\phi(0)/\varsigma_j^+$, so $|\partial F/\partial\mu_j^+|\le\phi(0)/\varsigma_j^+$ for every $\tilde\epsilon$ and every $s$; symmetrically for $\mu_j^-$. Since $\partial\mu_j^\pm/\partial\tilde\epsilon=\|c_j^\pm\|\tilde c_j^\pm/(\theta_j\kappa_j^\pm)$, the chain rule gives $\nabla_{\tilde\epsilon}F=\sum_{j\in\mathcal{H}_0}\sum_\pm\nu_j^\pm\tilde c_j^\pm$ with 
    \[
    |\nu_j^\pm|\le\frac{\phi(0)}{\varsigma_j^\pm}\cdot\frac{\|c_j^\pm\|}{\theta_j\kappa_j^\pm}=\phi(0)\theta_j\sqrt{\kappa_j^\pm}\cdot\frac{\sqrt{2a_j^\pm}}{\theta_j\kappa_j^\pm}=\phi(0)\sqrt{\frac{2a_j^\pm}{\kappa_j^\pm}}\le\phi(0),
    \]
    the factor $\theta_j$ cancelling and the last step following from $2a_j^\pm\le\kappa_j^\pm$. Hence $\|\nu\|^2\le2\phi(0)^2p_0$ and by \eqref{eq:Chat}, $\|\nabla_{\tilde\epsilon}F\|^2=\nu^\top\widetilde C^\top\widetilde C\nu\le\Lambda\|\nu\|^2\le2\phi(0)^2\Lambda p_0=:L^2$, so $F$ is $L$-Lipschitz on $\mathbb{R}^n$. The Gaussian concentration inequality for Lipschitz functions \citep[Theorem~5.6]{Boucheron2013} then gives
    \[
    \mathbb{P}_\mathbf{w}\big(|F-\mathbb{E}_\mathbf{w}F|>\varepsilon p_0/2\big)\le 2\exp\!\left(-\frac{\varepsilon^2p_0^2}{8L^2}\right)=2\exp\!\left(-\frac{\pi\varepsilon^2p_0}{8\Lambda}\right),
    \]
    using $\phi(0)^2=1/2\pi$. Combining the two bounds by a union bound gives \eqref{eq:conc}. For $D_{\sigma_0}^\pm$ the sums run over $\mathcal{H}_1$ and $\mu_j^\pm$ contains the additional term $a_j^\pm\beta_j^0/(\sigma_0\theta_j\kappa_j^\pm)$, which does not depend on $\tilde\epsilon$ and vanishes on differentiation. The gradient bound is unchanged with $\|\nu\|^2\le2\phi(0)^2p_1$.

    Let $f_j^0$ denote the density of $\widetilde M_j(\sigma_0)$ given $(X,w)$. We next show that this density is bounded by $\phi(0)\sqrt{a_j^{\max}/a_j^\star}$, so that
    \[
        L_{\sigma_0}:=\sup_{j,s}f^0_j(s)    \le\phi(0)(1+o(1))
    \]
    under Assumption~\ref{as:spectral}, by \eqref{eq:balance}. Given $X$ and $\mathbf{w}$, the variables $B_j^+$ and $B_j^-$ are independent, by $\mathrm{Cov}(g_j^+,g_j^-)=\sigma_0^2(c_j^+)^\top(c_j^-)=0$ from Lemma~\ref{th:1} and $z_j^+\perp\!\!\!\perp z_j^-$, with variances
    \[
    (\tilde{\kappa}_j^\pm)^2=\frac{2a_j^\pm}{\theta_j^2(\kappa_j^\pm)^2}+\frac{1}{\theta_j^2\kappa_j^\pm}=\frac{4a_j^\pm+\tau^{-2}}{\theta_j^2(\kappa_j^\pm)^2}.
    \]
    For $s>0$, differentiating the single-$j$ term of \eqref{eq:ap_F} in $s$, with $\tilde{\kappa}_j^\pm$ in place of $\varsigma_j^\pm$ since $B_j^\pm$ has unconditional standard deviation $\tilde{\kappa}_j^\pm$, and changing sign,
    \[
    f_j^0(s)=\frac{\phi(A_j^+)\bar\Phi(A_j^-)+\phi(\bar A_j^+)\Phi(\bar A_j^-)}{2\tilde{\kappa}_j^+}+\frac{\phi(A_j^-)\bar\Phi(A_j^+)+\phi(\bar A_j^-)\Phi(\bar A_j^+)}{2\tilde{\kappa}_j^-}.
    \]
    Bounding each $\phi$ by $\phi(0)$ and noting that $\bar A_j^\pm<A_j^\pm$, so that $\bar\Phi(A_j^\pm)+\Phi(\bar A_j^\pm)\le1$,
    \[
    f_j^0(s)\le\frac{\phi(0)}{2}\left(\frac{1}{\tilde{\kappa}_j^+}+\frac{1}{\tilde{\kappa}_j^-}\right).
    \]
    By \eqref{eq:ap_theta}, $\theta_j^2\le1/\kappa_j^{\min}$, and $4a_j^\pm+\tau^{-2}\ge2a_j^\pm+\tau^{-2}=\kappa_j^\pm$, so 
    \[
    \frac{1}{\tilde{\kappa}_j^\pm}=\frac{\theta_j\kappa_j^\pm}{\sqrt{4a_j^\pm+\tau^{-2}}}\le\frac{\kappa_j^\pm}{\sqrt{\kappa_j^{\min}\kappa_j^\pm}}=\sqrt{\frac{\kappa_j^\pm}{\kappa_j^{\min}}}\le\sqrt{\frac{a_j^{\max}}{a_j^\star}},
    \]
    the last step by $\kappa_j^{\max}/{\kappa_j^{\min}}\le a_j^{\max}/a_j^\star=(1+|\gamma_j|)/(1-|\gamma_j|)$. Hence $f_j^0\le\phi(0)\sqrt{a_j^{\max}/a_j^\star}$, and \eqref{eq:balance} gives $L_{\sigma_0}\le\phi(0)(1+o(1))$. The case $s<0$ is symmetric and $s=0$ is a single point.

    We now show that for every $s\in\mathbb{R}$ and  $\varepsilon>8L_{\sigma_0}\Delta_p$,
    \begin{equation}\label{eq:transfer}
        \mathbb{P}_\mathbf{w}\Big(\big|\bar V^+(s)-G_p(s)\big|>\varepsilon\Big)\le 4e^{-\varepsilon^2p_0/8}+4\exp\!\Big(-\frac{\pi\varepsilon^2p_0}{32\Lambda}\Big)+Cp^{-K}.
    \end{equation}
    On the coupling event $\mathcal{A}_p$ of Lemma~\ref{lem:5}, if $\widetilde M_j(\sigma_0)\ge s+\Delta_p$ then $\widetilde M_j\ge s$ and if $\widetilde M_j\ge s$ then $\widetilde M_j(\sigma_0)\ge s-\Delta_p$, so 
    \begin{equation*}
        V_{\sigma_0}^+(s+\Delta_p)\le V^+(s)\le V_{\sigma_0}^+(s-\Delta_p)\qquad\text{on }\mathcal{A}_p.
    \end{equation*}
    Since $u\mapsto\mathbb{E}_\mathbf{w}V_{\sigma_0}^+(u)=\sum_{j\in\mathcal{H}_0}\mathbb{P}_\mathbf{w}(\widetilde M_j(\sigma_0)\ge u)$ has derivative bounded by $p_0L_{\sigma_0}$ in absolute value,
    \begin{equation*}
        \big|\mathbb{E}_\mathbf{w}V_{\sigma_0}^+(s\mp\Delta_p)-\mathbb{E}_\mathbf{w}V_{\sigma_0}^+(s)\big|\le p_0L_{\sigma_0}\Delta_p.
    \end{equation*}
    If $\mathbf{1}(\widetilde M_j\ge s)\neq\mathbf{1}(\widetilde M_j(\sigma_0)\ge s)$ then $s$ lies weakly between the two values, so $|\widetilde M_j(\sigma_0)-s|\le\Delta_p$ on $\mathcal{A}_p$. Hence $|\mathbb{P}_\mathbf{w}(\widetilde M_j\ge s)-\mathbb{P}_\mathbf{w}(\widetilde M_j(\sigma_0)\ge s)|\le\mathbb{P}_\mathbf{w}(|\widetilde M_j(\sigma_0)-s|\le\Delta_p)+\mathbb{P}_\mathbf{w}(\mathcal{A}_p^c)\le2L_{\sigma_0}\Delta_p+Cp^{-K}$ by Lemma~\ref{lem:5}, and summing over $\mathcal{H}_0$,
    \begin{equation*}
        \big|p_0G_p(s)-\mathbb{E}_\mathbf{w}V_{\sigma_0}^+(s)\big|\le p_0\big(2L_{\sigma_0}\Delta_p+Cp^{-K}\big),
    \end{equation*}
    using $\mathbb{E}_\mathbf{w}V^+(s)=p_0G_p(s)$ from \eqref{eq:Gp}. Fix $\varepsilon>8L_{\sigma_0}\Delta_p$ and let $\mathcal{E}$ be the event that $|V_{\sigma_0}^+(s\mp\Delta_p)-\mathbb{E}_\mathbf{w}V_{\sigma_0}^+(s\mp\Delta_p)|\le\varepsilon p_0/2$ for both signs; by \eqref{eq:conc} and a union bound,
    $\mathbb{P}_\mathbf{w}(\mathcal{E}^c)\le4e^{-\varepsilon^2p_0/8}+4\exp\{-\pi\varepsilon^2p_0/(32\Lambda)\}$. On $\mathcal{A}_p\cap\mathcal{E}$, combining the three equations above,
    \[
    \big|V^+(s)-p_0G_p(s)\big|\le\frac{\varepsilon p_0}{2}+p_0L_{\sigma_0}\Delta_p+p_0\big(2L_{\sigma_0}\Delta_p+Cp^{-K}\big)\le\varepsilon p_0
    \]
    for sufficiently large $p$, since $3L_{\sigma_0}\Delta_p<3\varepsilon/8$ and $Cp^{-K}\to0$. Therefore $\mathbb{P}_\mathbf{w}(|\bar V^+(s)-G_p(s)|>\varepsilon)\le\mathbb{P}_\mathbf{w}(\mathcal{E}^c)+\mathbb{P}_\mathbf{w}(\mathcal{A}_p^c)$, which is \eqref{eq:transfer}. 

    Fix $\varepsilon>16L_{\sigma_0}\Delta_p$. The map $G_p$ is continuous, since each $\widetilde M_j$ has a density, non-increasing, with $G_p(-\infty)=1$ and $G_p(+\infty)=0$. By the intermediate value theorem there exists $-\infty=s_0<s_1<\dots<s_N=+\infty$, with $G_p(s_{k-1})-G_p(s_k)\le\varepsilon/2$ for each $k$ and $N\le\lceil2/\varepsilon\rceil$. Since $\bar V^+$ and $G_p$ are both non-increasing, for $s\in[s_{k-1},s_k]$,
    \[
    \bar V^+(s)-G_p(s)\le\bar V^+(s_{k-1})-G_p(s_k)\le\big\{\bar V^+(s_{k-1})-G_p(s_{k-1})\big\}+\varepsilon/2,
    \]
    \[
    \bar V^+(s)-G_p(s)\ge\bar V^+(s_k)-G_p(s_{k-1})\ge\big\{\bar V^+(s_k)-G_p(s_k)\}-\varepsilon/2.
    \]
    Hence $\sup_s|\bar V^+(s)-G_p(s)|\le\max_{1\le k\le N-1}|\bar V^+(s_k)-G_p(s_k)|+\varepsilon/2$, the endpoints $s_0,s_N$ contributing zero since $\bar V^+$ and $G_p$ agree there. Applying \eqref{eq:transfer} at level $\varepsilon/2$, which is admissible because $\varepsilon/2>8L_{\sigma_0}\Delta_p$, and a union bound over the at most $\lceil2/\varepsilon\rceil$ interior grid points, gives \eqref{eq:uniform}. The argument for $\bar V^-$ and $\bar D^\pm$ is identical, with $p_1$ in place of $p_0$ for the signal count.
    
\end{proof}

\begin{proof}[Proof of Proposition~\ref{prop:spectral}]
    By \eqref{eq:c_def}, $c_j^\pm=\sqrt{\alpha_j}(\tilde q_j\pm\tilde w_j)$ with $\|c_j^\pm\|^2=2\alpha_j(1\pm\gamma_j)$, so $\tilde c_j^\pm=(\tilde q_j\pm\tilde w_j)/\sqrt{2(1\pm\gamma_j)}$. Substituting $\tilde w_j=\sqrt{1-\gamma_j^2}\,\tilde u_j+\gamma_j\tilde q_j$ from Lemma~\ref{lem:2} and using $\sqrt{1-\gamma_j^2}/\sqrt{2(1\pm\gamma_j)}=\sqrt{(1\mp\gamma_j)/2}$ and $(1\pm\gamma_j)/\sqrt{2(1\pm\gamma_j)}=\sqrt{(1\pm\gamma_j)/2}$,
    \begin{equation*}
        \tilde c_j^\pm=b_j^\pm\tilde q_j\pm b_j^\mp\tilde u_j, \qquad b_j^\pm=\sqrt{(1\pm\gamma_j)/2}.
    \end{equation*}
    Ordering the columns of $\widetilde C$ as $(1,+),(1,-),\dots,(p,+),(p,-)$ and defining $\Gamma_Q,\Gamma_U\in\mathbb{R}^{p\times 2p}$ by placing $b_j^\pm$, respectively $\pm b_j^\mp$, in row $j$ of column $(j,\pm)$ and zero elsewhere, we obtain $\widetilde C= Q\Gamma_Q+\widetilde U\Gamma_U$.

    Each row of $\Gamma_Q$ has exactly two non-zero entries and distinct rows have disjoint supports, so $\Gamma_Q\Gamma_Q^\top$ is diagonal with entries $(b_j^+)^2+(b_j^-)^2=(1+\gamma_j)/2+(1-\gamma_j)/2=1$, thus $\Gamma_Q\Gamma_Q^\top=I_p$ and every singular value of $\Gamma_Q$ equals one. The same computation gives $\Gamma_U\Gamma_U^\top=I_p$. Both identities hold for every $\gamma_j\in[-1,1]$ and the triangle inequality gives $\sqrt{\Lambda}=\|\widetilde C\|_{\mathrm{op}}\le\|Q\|_{\mathrm{op}}\|\Gamma_Q\|_{\mathrm{op}}+\|\widetilde U\|_{\mathrm{op}}\|\Gamma_U\|_{\mathrm{op}}$.

    By Lemma~\ref{lem:1}, $\| Q\|_{\mathrm{op}}^2=\lambda_{\max}(\widetilde\Omega)$. For any unit $x$, $x^\top \widetilde\Omega x\le\lambda_{\max}(\Omega)\| D_\Omega^{-1/2}x\|^2\le\lambda_{\max}(\Omega)/\min_j\Omega_{jj}$, and $\lambda_{\max}(\Omega)=1/\lambda_{\min}(X^\top X)$ while $\Omega_{jj}=\iota_j^\top(X^\top X)^{-1}\iota_j\ge1/\lambda_{\max}(X^\top X)$, hence $\lambda_{\max}(\widetilde\Omega)\le\mathrm{cond}(X^\top X)$.

    Condition on $X$. By Lemma~\ref{lem:2} the $\tilde u_j$ are i.i.d.\ uniform on the unit sphere of the fixed $(n-p)$-dimensional subspace $\mathrm{col}(H)$. Fixing an orthonormal basis, write $\tilde u_j=\tilde g_j/\|\tilde g_j\|$ with $\tilde g_j\sim N(0,I_{n-p})$ i.i.d., so $\widetilde U=\widetilde GN^{-1}$ with $N=\mathrm{diag}(\|\tilde g_j\|)$ and $\|\widetilde U\|_{\mathrm{op}}\le\|\widetilde G\|_{\mathrm{op}}/\min\|\tilde g_j\|$. By Theorem~2.13 of \citet{Davidson2001}, applied to $\widetilde G$ after rescaling the entries to unit variance, $\mathbb{P}(\|\widetilde G\|_{\mathrm{op}}>\sqrt{n-p}+\sqrt p+u)\le e^{-u^2/2}$, and by Lemma~1 of \citet{LaurentMassart2000-Adapt}, $\mathbb{P}(\|\tilde g_j\|^2\le n-p-2\sqrt{(n-p)x})\le e^{-x}$, so taking $x=K\log p$ and a union bound gives $\min_j\|\tilde g_j\|^2\ge n-p-2\sqrt{(n-p)K\log p}\ge (n-p)/2$ with probability at least $1-p^{1-K}$ whenever $n-p\ge16K\log p$. Taking $u=\sqrt{2K\log p}$ and combining,
    \[
    \|\widetilde U\|_{\mathrm{op}}\le\frac{\sqrt{n-p}+\sqrt p+\sqrt{2K\log p}}{\sqrt{(n-p)/2}}=\sqrt{2}\left(1+\sqrt{\frac {p}{n-p}}+\sqrt{\frac{2K\log p}{n-p}}\right)
    \]
    with probability at least $1-e^{-K\log p}-p^{1-K}\ge1-2p^{1-K}$.
\end{proof}

\begin{proof}[Proof of Lemma~\ref{lem:9}]
    Since $\pi_0+\pi_1=1$,
    \[
    G_R(s_q)\ge \min\big\{G_p(s_q),G^+_{1,p}(s_q)\big\}\longrightarrow\min\big\{G_0(s_q),G^+_1(s_q)\big\}=:G_{\min}>0
    \]
    by (C2), (C3) and (C4), so $G_R(s_q)\ge3 G_{\min}/4$ for sufficiently large $p$. The map $(\pi_0,a,b,c,d)\mapsto\{\pi_0a+(1-\pi_0)b\}/\{\pi_0c+(1-\pi_0)d\}$ is uniformly continuous on the set where $\pi_0\in[0,1]$ and the denominator is at least $3G_{\min}/4$. Applying it with $(a,b,c,d)=(G_p,G^-_{1,p},G_p,G^+_{1,p})$ and with the corresponding limits, (C2) gives
    \[
    \frac{G_A(s_q)}{G_R(s_q)}-\frac{\pi_0G_0(s_q)+\pi_1G^-_1(s_q)}{\pi_0G_0(s_q)+\pi_1G^+_1(s_q)}\longrightarrow0.
    \]
    By (C3) the second term is at most $q-2\delta$ for some $\delta>0$ and all $p$ large, hence $G_A(s_q)\le(q-\delta)G_R(s_q)$ for all sufficiently large $p$.

    Set $\mathcal{B}_p=\{\eta_p\le\varepsilon_0\}$ with $\varepsilon_0=\tfrac{1}{4}G_{\min}\min\{1,\delta/(1+q)\}$. Then $\mathbb{P}_\mathbf{w}(\mathcal{B}_p)\to1$, since $\eta_p=o_p(1)$, as shown in Section~\ref{sec:fdr}.

     On $\mathcal{B}_p$, \eqref{eq:ARdev} gives $\bar R(s_q)\ge G_R(s_q)-\varepsilon_0\ge3G_{\min}/4-G_{\min}/4=G_{\min}/2$, so $R(s_q)\ge G_{\min}p/2\ge1$ for sufficiently large $p$ and $R(s_q)\vee1=R(s_q)$. Hence
    \[
    \widehat{\mathrm{FDP}}(s_q)=\frac{p^{-1}+\bar A(s_q)}{\bar R(s_q)}\le\frac{p^{-1}+G_A(s_q)+\eta_p}{G_R(s_q)-\eta_p}\le q \quad \text{whenever}
    \quad  p^{-1}+(1+q)\eta_p\le\delta\,G_R(s_q),
    \]
    since $q\,G_R(s_q)-G_A(s_q)\ge\delta G_R(s_q)$. The right-hand side is at least $3\delta G_{\min}/4$, while on $\mathcal{B}_p$, the left-hand side is at most $p^{-1}+\delta G_{\min}/4$, which is smaller for sufficiently large $p$.  Therefore $s_q$ lies in the set defining the threshold $\tau_q^{(t)}$ and $\tau_q^{(t)}\le s_q$.

    Write $\tau_q=\tau_q^{(t)}$ and $\widehat{\mathrm{FDP}}(\tau_q+)=\lim_{s\downarrow\tau_q}\widehat{\mathrm{FDP}}(s)$. Since $A$ and $R$ are step functions with jumps only in the finite set $\{|\widetilde M_j^{(t)}|\}_{j=1}^{p}$, any sequence $s_n\downarrow\tau_q$ with $\widehat{\mathrm{FDP}}(s_n)\le q$, eventually lies in one interval of constancy, so $\widehat{\mathrm{FDP}}(\tau_q+)\le q$. Since the $\widetilde M_j^{(t)}$ are continuous random variables, almost surely their absolute values are distinct, so at most one index satisfies $|\widetilde M_j^{(t)}|=\tau_q$. If none does, or if that index has $\widetilde M_j^{(t)}=\tau_q$, then $A(\tau_q)=A(\tau_q+)$ and $R(\tau_q)\ge R(\tau_q+)$, giving $\widehat{\mathrm{FDP}}(\tau_q)\le q$. If it has $\widetilde M_j^{(t)}=-\tau_q$, then $A(\tau_q)=A(\tau_q+)+1$ and $R(\tau_q)= R(\tau_q+)$, so $\widehat{\mathrm{FDP}}(\tau_q)\le q+1/R(\tau_q)$, the denominator being at least $R(s_q)\ge 1$ by the previous step and monotonicity. 
\end{proof}

\begin{proof}[Proof of Theorem~\ref{th:2}]
    We abbreviate $\tau_q=\tau_q^{(t)}$ throughout and work on $\mathcal{B}_p$ of Lemma~\ref{lem:9}, where $\tau_q\le s_q$ and $\bar R$ is non-increasing, $\bar R(\tau_q)\ge\bar R(s_q)\ge G_{\min}/2$ so $R(\tau_q)\ge G_{\min}p/2\ge1$. Using $A=V^-+D^-\ge V^-$ and then \eqref{eq:diff}, 
    \[
    \mathrm{FDP}(\tau_q)-\widehat{\mathrm{FDP}}(\tau_q)=\frac{V^+(\tau_q)-1-A(\tau_q)}{R(\tau_q)}\le\frac{V^+(\tau_q)-V^-(\tau_q)}{R(\tau_q)}=\frac{\pi_0\left(\bar V^+(\tau_q)-\bar V^-(\tau_q)\right)}{\bar R(\tau_q)}
    \le\frac{4\eta_p^0}{G_{\min}},
    \]
    using $\pi_0\le1$. Adding the threshold bound of Lemma~\ref{lem:9},
    \[
    \mathrm{FDP}(\tau_q)\ \le\ q+\frac{2}{G_{\min} p}+\frac{4\eta_p^0}{G_{\min}}\ =\ q+o_p(1),
    \]
    since $\eta_p=o_p(1)$ and $\mathbb{P}_\mathbf{w}(\mathcal{B}_p)\to1$. For the second claim, let $\Theta_p$ be the sum of the two error terms, so $\mathrm{FDP}(\tau_q)\le q+\Theta_p$ on $\mathcal{B}_p$ with $\Theta_p=o_p(1)$. Since $\mathrm{FDP}\in[0,1]$ 
    \[
    \mathbb{E}_\mathbf{w}\big[\mathrm{FDP}(\tau_q)\big]\le 
    \mathbb{E}_\mathbf{w}\big[\mathrm{FDP}(\tau_q)\mathbf{1}_{\mathcal{B}_p}\big]+\mathbb{P}_\mathbf{w}(\mathcal{B}_p^c)\le 
    q+\mathbb{E}_\mathbf{w}\big[\min\{\Theta_p,1\}\big]+\mathbb{P}_\mathbf{w}(\mathcal{B}_p^c),
    \]
    the middle term tending to zero because a uniformly bounded sequence converging in probability converges in $L^1$, and the last by Lemma~\ref{lem:9}. 
\end{proof}

\begin{proof}[Proof of Corollary~\ref{cor:ratio}]
    On $\mathcal{B}_p$ the threshold satisfies $\tau_q^{(t)}\le s_q$, and $V^-$ is non-increasing, so by \eqref{eq:etazeta}
    \[A(\tau_q^{(t)})\;\ge\;V^-(\tau_q^{(t)})\;\ge\;V^-(s_q)\;\ge\;p_0\big(G_p(s_q)-\eta_p^0\big).\]
    By (C2) at $s_q$ and (C4), $G_p(s_q)\ge3G_0(s_q)/4$ for sufficiently large $p$, so on $\mathcal{D}_p$ this is at least $p_0G_0(s_q)/2=c_Ap_0$. Moreover, by (\ref{eq:diff}) and $A\ge V^-$,
    \[\frac{V^+(\tau_q^{(t)})}{1+A(\tau_q^{(t)})}\;\le\;\frac{V^-(\tau_q^{(t)})+2\eta_p^0p_0}{1+A(\tau_q^{(t)})}\;\le\;1+\frac{2\eta_p^0p_0}{1+A(\tau_q^{(t)})}\;\le\;1+\frac{2\eta_p^0}{c_A}.\]
    The probability bound follows from Lemma~\ref{lem:9} and \eqref{eq:uniform}.
\end{proof}

\begin{proof}[Proof of Lemma~\ref{lem:10}]
    We have $\mathcal{D}_p=\mathcal{B}_p\cap\{\eta_p^0\le G_0(s_q)/4\}$ with $\mathcal{B}_p=\{\eta_p^0+\eta_p^1\le\varepsilon_0\}$, so 
    \[
    \mathcal{D}_p^c\subseteq \{\eta_p^0>\varepsilon_0/2\}\cup\{\eta_p^1>\varepsilon_0/2\}\cup\{\eta_p^0>G_0(s_q)/4\},
    \]
    a union of three events, each a deviation of $\eta_p^0$ or $\eta_p^1$ above a level not depending on $p$. Applying \eqref{eq:uniform} at those levels, and noting that $\liminf p_1/p_0>0$ under (C3)--(C4), gives $\mathbb{P}_\mathbf{w}(\mathcal{D}_p^c)\le C\exp\{-cp_0/(1\vee\Lambda)\}+Cp^{-K}$. Multiplying by $p_0$, the first term vanishes because $\Lambda=o(p_0/\log p_0)$ by Assumption~\ref{as:spectral}, so $p_0/(1\vee\Lambda)\gg\log p_0$, while the second vanishes because $p_0\le p$ and $K\ge 2$.
\end{proof}

\begin{proof}[Proof of Theorem~\ref{th:3}]
    Write $\Upsilon_p=V^+(\tau_q^{(t)})/\big(1+A(\tau_q^{(t)})\big)$, so that $\sum_{j\in\mathcal{H}_0}\mathbb{E}_{\mathbf{w}}[e_j^{(t)}]=p\,\mathbb{E}_{\mathbf{w}}[\Upsilon_p]$. On $\mathcal{D}_p$, Corollary~\ref{cor:ratio} gives $A(\tau_q^{(t)})\ge c_Ap_0$ and hence the deterministic bound $\Upsilon_p\le p_0/(c_Ap_0)=1/c_A$, together with $\Upsilon_p\le1+2\eta_p^0/c_A$. Therefore
    \[\mathbb{E}_{\mathbf{w}}\big[\Upsilon_p\mathbf{1}_{\mathcal{D}_p}\big]\;\le\;\mathbb{E}_{\mathbf{w}}\!\left[\min\{1+2\eta_p^0/c_A,\;1/c_A\}\right]\;\longrightarrow\;1,\]
    since the integrand is bounded by the fixed constant $1/c_A$ and converges to $1$ in probability, and a uniformly bounded sequence converging in probability converges in $L^1$. Off $\mathcal{D}_p$ we use $\Upsilon_p\le V^+\le p_0$, so $\mathbb{E}_{\mathbf{w}}[\Upsilon_p\mathbf{1}_{\mathcal{D}_p^c}]\le p_0\mathbb{P}_{\mathbf{w}}(\mathcal{D}_p^c)\to0$ by Lemma~\ref{lem:10}. Taking $\delta_p$ to be the sum of the two error terms gives $\mathbb{E}_{\mathbf{w}}[\Upsilon_p]\le1+\delta_p$ with $\delta_p\to0$. Since the iterates are identically distributed, $\delta_p$ does not depend on $t$, and the second display follows by linearity of the expectation.
\end{proof}

\begin{proof}[Proof of Theorem~\ref{th:4}]
    Let $\delta_p\to0$ be the sequence of Theorem~\ref{th:3} and set $\tilde e_j=\bar e_j/(1+\delta_p)$, so that the $\tilde e_j$ are non-negative with $\sum_{j\in\mathcal{H}_0}\mathbb{E}_\mathbf{w}[\tilde e_j]\le p$. They therefore satisfy the requirement of the e-BH procedure in the relaxed form \citep{Wang2022False-Discovery,Ren2024Derandomised-kn}, which places no condition on the dependence among them.

    Since $\tilde e_j$ is a common positive multiple of $\bar e_j$, applying e-BH to $\{\tilde e_j\}$ at level $q^{\star\prime}=q^\star(1+\delta_p)$ returns exactly $\widehat{\mathcal{S}}$, since the order statistics satisfy $\tilde e_{(k)}=\bar e_{(k)}/(1+\delta_p)$, so $\tilde e_{(k)}\ge p/(q^{\star\prime} k)$ if and only if $\bar e_{(k)}\ge p/(q^\star k)$, hence the two procedures share same $\hat k$ and the same selected set. The e-BH guarantee at level $q^{\star\prime}$ gives
    \[
    \mathbb{E}_{\mathbf{w}}\!\left[\frac{|\mathcal{H}_0\cap\widehat{\mathcal{S}}|}{|\widehat{\mathcal{S}}|\vee1}\right]\;\le\;q^{\star\prime}=q^\star(1+\delta_p),
    \]
    which is the stated bound since $\delta_p\to0$.
\end{proof}

\begin{proof}[Proof of Corollary~\ref{cor:uncond}]
    The FDP lies in $[0,1]$, so on the intersection of the events of Lemma~\ref{lem:4} and Proposition~\ref{prop:spectral}, of probability at least $1-4p^{1-K}$, the bound of Theorem~\ref{th:4} holds with $\delta_p$ replaced by $\bar\delta_p$, and off it the proportion is at most one. Taking expectations over $\mathbf{w}$ and applying the tower property gives the claim.
\end{proof}

\begin{proof}[Proof of Proposition~\ref{prop:hd}]
    Condition on $\widehat{J}$ and work on $\mathcal{F}_p$. By (S1), $\beta_j^0=0$ for every $j\notin\widehat{J}$, so
    \[
    y^{(2)}=X^{(2)}_{\widehat{J}}\beta^0_{\widehat{J}}+\epsilon^{(2)},
    \]
    a linear model in $\hat p<n_2$ variables whose null set is exactly $\mathcal{H}_0\cap\widehat{J}$.

    By (S2) the design has full column rank with $\hat r\ge2$, so Lemma~\ref{th:1} and Lemmas~\ref{lem:1}--\ref{lem:2}, which are statements about an arbitrary full-rank design with more rows than columns, hold for $X^{(2)}_{\widehat{J}}$. Lemma~\ref{lem:3} is an algebraic identity. By the displayed model above and the exogeneity established in Section~\ref{sec:high_dim}, both hypotheses of Lemma~\ref{lemma1} are met for every $j\in\mathcal{H}_0\cap\widehat{J}$, so Lemma~\ref{lemma1} and Corollary~\ref{cor:all_perturbations} hold on the screened problem.

    The results of Section~\ref{sec:toolkit} consume only Assumptions~\ref{as:design}--\ref{as:regime} together with the symmetry just established, so by (S3) Lemmas~\ref{lem:4}--\ref{lem:5} and Proposition~\ref{prop:conc} hold with $\hat p_0,\hat p_1,\widehat\Lambda,\hat r$ in place of $p_0,p_1,\Lambda,r$. Conditions (C1)--(C4) for the screened problem then give Lemma~\ref{lem:9} and Theorem~\ref{th:2}, and with Lemma~\ref{lem:10}, Theorems~\ref{th:3} and~\ref{th:4}, and by (S3), Corollary~\ref{cor:uncond}, all conditionally on $\widehat{J}$.

    For $j\notin\widehat{J}$ we have $\widetilde M_j^{(t)}=0$, which is neither $\ge s$ nor $\le-s$ for any $s>0$, so such coordinates enter neither $R$ nor $A$ and have $e_j^{(t)}=0$. Finally the FDP lies in $[0,1]$ and the bound holds with the same constants for every realisation of $\widehat{J}$ in $\mathcal{F}_p$, so the tower property gives the unconditional statement at an additive cost of $\mathbb{P}(\mathcal{F}_p^c)\to0$.
\end{proof}

\begin{proof}[Proof of Lemma~\ref{lem:power-A}]
    $f'(a)=\tau^{-2}/(2a+\tau^{-2})^2\ge0$, so $f$ is non-decreasing. Combined with the identity $x+y-|x-y|=2\min(x,y)$ applied to $x=f(a_j^+), y=f(a_j^-)$, this gives $\lambda_u-|\lambda_\Delta|=2f(\min(a_j^+,a_j^-))=2f(a_j^\star)$, positive since $a_j^\star>0$. Then $|u_j|\ge|\beta_j^0|\lambda_u-|E_u|$ because $\lambda_u>0$, and $|\Delta_j|\le|\beta_j^0||\lambda_\Delta|+|E_\Delta|$. Subtracting gives the result.
\end{proof}

\begin{proof}[Proof of Lemma~\ref{lem:power-std}]
    By Lemma~\ref{lem:power-A} and $\hat s_j=\sigma_j\theta_j$, $\widetilde M_j\ge2f(a_j^\star)|\beta_j^0|/(\sigma_j\theta_j)-\Xi_j$. For the first term, $2f(a_j^\star)=2a_j^\star/\kappa_j^\star$ with $\kappa_j^\star=2a_j^\star+\tau^{-2}=\kappa_j^{\min}$, and $\theta_j\le(\kappa_j^{\min})^{-1/2}$ by \eqref{eq:ap_theta}, so
    \[
    \frac{2f(a_j^\star)|\beta_j^0|}{\sigma_j\theta_j}\ge \frac{2a_j^\star|\beta_j^0|}{\sigma_j\sqrt{\kappa_j^\star}}=\frac{\sigma_0}{\sigma_j}\psi_j\frac{2\sqrt{a_j^\star}}{\sqrt{2a_j^\star+\tau^{-2}}}.
    \]
    By Assumption~\ref{as:spectral}, $|\gamma_j|\le1/2$ for sufficiently large $p$, so $a_j^\star=\alpha_j(1-|\gamma_j|)\ge\min_k\alpha_k/2$, and Assumption~\ref{as:design} gives $\tau^{-2}\le{c_v}\min_k\alpha_k\le2{c_v}a_j^\star$. Hence $2a_j^\star+\tau^{-2}\le2(1+{c_v})a_j^\star$ and the last factor is at least $\sqrt{2/(1+{c_v})}=c_\psi$. On the event of \eqref{eq:ap_rho}, $|\sigma_0/\sigma_j-1|\le C\omega_p$, so $\sigma_0/\sigma_j\ge1-C\omega_p$ which gives \eqref{eq:std-sep}. As $\tau^{-2}\to0$ we may take ${c_v}\to0$, giving $c_\psi\to\sqrt2$.
\end{proof}

\begin{proof}[Proof of Theorem~\ref{th:5}]
   Recall $\kappa_j^\pm\ge2a_j^\star$ and $\theta_j\ge(2\kappa_j^{\max})^{-1/2}$ by \eqref{eq:ap_theta},
    \[
    \Xi_j\le\frac{2\sqrt{2\kappa_j^{\max}}}{\sigma_j}\left(\frac{|g_j^+|}{\kappa_j^+}+\frac{|g_j^-|}{\kappa_j^-}\right)+2\sqrt{2\kappa_j^{\max}}\left(\frac{|z_j^+|}{\sqrt{\kappa_j^+}}+\frac{|z_j^-|}{\sqrt{\kappa_j^-}}\right).
    \]
    Using $\kappa_j^\pm\ge\|c_j^\pm\|^2=2a_j^\pm$ and $\kappa_j^{\max}/\kappa_j^{\min}\le1+o(1)$, together with $\sigma_0/\sigma_j\le1+C\omega_p$ on the event of \eqref{eq:ap_rho}, there is an absolute constant $C_1$ with 
    \[
    \Xi_j\le C_1\max\Big\{\big|g_j^+\big|\big/\sigma_0\sqrt{2a_j^+},\ \big|g_j^-\big|\big/\sigma_0\sqrt{2a_j^-},\ |z_j^+|,\ |z_j^-| \Big\}
    \]
    under Assumption~\ref{as:spectral} and on the event of \eqref{eq:ap_rho}. Under $\mathbb{P}_\mathbf{w}$ the vectors $c_j^\pm$ are fixed and $\epsilon\sim N(0,\sigma_0^2I_n)$, so by \eqref{eq:d-comp}, $g_j^\pm=(c_j^\pm)^\top\epsilon\sim N(0,2\sigma_0^2a_j^\pm)$ and hence all four variables above are standard normal. A union bound over four two-sided Gaussian tails gives
    \begin{equation}\label{eq:Xi-tail}
        \mathbb{P}_\mathbf{w}\big(\Xi_j>\eta\big)\le8\exp\!\big(-\eta^2/2C_1^2\big),\qquad\eta>0.
    \end{equation}

    Let $\mathcal{E}_p$ be the intersection of $\{\Xi_j\le m_p/2\}$ with the event of \eqref{eq:ap_rho}. On $\mathcal{E}_p$, \eqref{eq:std-sep} and (A1) give
    \[
    \widetilde M_j\ge c_\psi\psi_j-C\omega_p c_\psi\psi_j-\Xi_j\ge s_q+m_p-C\omega_p c_\psi\psi_{\max}-m_p/2=s_q+m_p/2-C\omega_p c_\psi\psi_{\max}.
    \]
    By Assumption~\ref{as:regime}, $\omega_p\psi_{\max}\to0$, so the subtracted term is eventually smaller than $m_p/2$, which is bounded below by a positive constant. Hence $\widetilde M_j>s_q$ on $\mathcal{E}_p$
    for all sufficiently large $p$. The correction does not scale with $m_p$, so the argument is unaffected by $m_p\to\infty$.

    On $\mathcal{E}_p\cap\mathcal{B}_p$ we have $\widetilde M_j>s_q\ge\tau_q^{(t)}$ by Lemma~\ref{lem:9}, that is $j\in\widehat{\mathcal{S}}^{(t)}$. By \eqref{eq:Xi-tail} at level $m_p/2$ and the tail bound in \eqref{eq:ap_rho},
    \[
    \mathbb{P}_\mathbf{w}\big(\mathcal{E}_p^c\big)\le 8\exp\!\big(-m_p^2/8C_1^2\big)+Cp^{-K},
    \]
    and adding $\mathbb{P}_\mathbf{w}(\mathcal{B}_p^c)$ gives the stated bound.
\end{proof}

\begin{proof}[Proof of Corollary~\ref{cor:power-global}]
     By \eqref{eq:Xi-tail} at level $m_p/2$ and a union bound over $j\in\mathcal{H}_1$,
     \[
     \mathbb{P}_\mathbf{w}\Big(\max_{j\in\mathcal{H}_1}\Xi_j>m_p/2\Big)\le8p_1\exp\!\big(-m_p^2/8C_1^2\big)\le 8p_1^{-K}\longrightarrow0.
     \]
     On the complement, intersected with the event of \eqref{eq:ap_rho} and $\mathcal{B}_p$, the threshold result of the proof of Theorem~\ref{th:5} applies simultaneously to every $j\in\mathcal{H}_1$, giving $\mathcal{H}_1\subseteq\widehat{\mathcal{S}}^{(t)}$. The second claim follows since the ratio is bounded by one and equals one on that event.
\end{proof}

\begin{proof}[Proof of Theorem~\ref{th:6}]
    For each $t$, Corollary~\ref{cor:power-global} exhibits an event $\mathcal{G}^{(t)}$ on which $\mathcal{H}_1\subseteq\widehat{\mathcal{S}}^{(t)}$, with
    \[
    \mathbb{P}_\mathbf{w}\big(\mathcal{G}^{(t)c}\big)\le8p_1^{-K}+Cp^{-K}+\mathbb{P}_\mathbf{w}\big(\mathcal{B}_p^c\big).
    \]
    The bound does not depend on $t$, since the iterates are identically distributed. By \eqref{eq:alpha-condition}, we may fix $\bar\varepsilon>0$ with $q^\star(1-q-q\bar\varepsilon)>q$. Let $\bar{\mathcal{D}}_p^{(t)}=\mathcal{D}_p^{(t)}\cap\{\eta_p^0\le c_A\bar\varepsilon/2\}$, with $\mathcal{D}_p^{(t)}$ and $c_A$ as in Corollary~\ref{cor:ratio} at iterate $t$. As in the proof of Lemma~\ref{lem:10}, the complement of $\bar{\mathcal{D}}_p^{(t)}$ is contained in a union of deviations of $\eta_p^0$ and $\eta_p^1$ above levels not depending on $p$, so \eqref{eq:uniform} gives $\mathbb{P}_\mathbf{w}(\bar{\mathcal{D}}_p^{(t)c})\le C\exp\{-c'p_0/(1\vee\Lambda)\}+Cp^{-K}$ for some $c'>0$ not depending on $t$. Since $\bar{\mathcal{D}}_p^{(t)}\subseteq\mathcal{B}_p$, the same bound holds for $\mathbb{P}_\mathbf{w}(\mathcal{B}_p^c)$.
    Let $\mathcal{E}=\bigcap_{t=1}^{T}\mathcal{G}^{(t)}$ and $\mathcal{D}=\bigcap_{t=1}^{T}\bar{\mathcal{D}}_p^{(t)}$.
    By a union bound over $t$ and $p_1\le p$,
    \begin{equation}\label{eq:E-union}
        \mathbb{P}_\mathbf{w}\big(\mathcal{E}^c\cup\mathcal{D}^c\big)\le C\,T\Big\{p_1^{-K}+\exp\Big(-\frac{c'p_0}{1\vee\Lambda}\Big)\Big\}\longrightarrow0
    \end{equation}
    by \eqref{eq:T-condition}.

    Work on $\mathcal{E}\cap\mathcal{D}$ and fix $t$. Since $\bar{\mathcal{D}}_p^{(t)}\subseteq\mathcal{D}_p^{(t)}$, Corollary~\ref{cor:ratio} gives 
    \begin{equation}\label{eq:event-D}
       \frac{V^+\bigl(\tau_q^{(t)}\bigr)}{1+A\bigl(\tau_q^{(t)}\bigr)}\le1+\frac{2\eta_p^0}{c_A}\le1+\bar\varepsilon.
    \end{equation}
    Since $\bar{\mathcal{D}}_p^{(t)}\subseteq\mathcal{B}_p$, Lemma~\ref{lem:9} gives $\tau_q^{(t)}\le s_q$ and
    \begin{equation}\label{eq:fdphat-at-tau}
        \widehat{\mathrm{FDP}}\big(\tau_q^{(t)}\big)=\frac{1+A\bigl(\tau_q^{(t)}\bigr)}{R\bigl(\tau_q^{(t)}\bigr)\vee1}\le q+\frac{1}{R\big(\tau_q^{(t)}\big)}.
    \end{equation}
    On $\mathcal{E}$ every active variable satisfies $\widetilde M_j^{(t)}\ge\tau_q^{(t)}$, so 
    \begin{equation*}
        R\bigl(\tau_q^{(t)}\bigr)=D^+\big(\tau_q^{(t)}\big)+V^+\bigl(\tau_q^{(t)}\bigr)=p_1+V^+\big(\tau_q^{(t)}\big)\ge p_1\ge1,
    \end{equation*}
    and \eqref{eq:fdphat-at-tau} becomes $1+A\le q\,p_1+qV^++1$, writing $A=A\bigl(\tau_q^{(t)}\bigr)$ and $V^+=V^+\bigl(\tau_q^{(t)}\bigr)$.
    Applying \eqref{eq:event-D} to $V^+$,
    \[
        1+A\le q\,p_1+q(1+\bar\varepsilon)(1+A)+1, \quad\text{that is}\quad (1+A)(1-q-q\bar\varepsilon)\le q\,p_1+1.
    \]
    By the choice of $\bar\varepsilon$, $1-q-q\bar\varepsilon>q/q^\star>0$. Every $j\in \mathcal{H}_1$ satisfies $\mathbf{1}\{\widetilde M_j^{(t)}\ge\tau_q^{(t)}\}=1$ for every $t$, so
    \begin{equation*}
        e_j^{(t)}=\frac{p}{1+A\bigl(\tau_q^{(t)}\bigr)}\ge\frac{p\,(1-q-q\bar\varepsilon)}{q\,p_1\,+1}, \qquad t=1,\dots,T,
    \end{equation*}
    and averaging over $t$ preserves the bound for $\bar e_j$. Since $q^\star(1-q-q\bar\varepsilon)>q$ and $p_1\to\infty$ by (C1), for all sufficiently large $p$, $q^\star(1-q-q\bar\varepsilon)\,p_1\ge q\,p_1+1$, equivalently
    \begin{equation*}
        \frac{p\,(1-q-q\bar\varepsilon)}{q\,p_1\,+1}\ge\frac{p}{q^\star\,p_1}.
    \end{equation*}
    
    Combining with the bound on the averaged e-values above, on $\mathcal{E}\cap\mathcal{D}$ all $p_1$ active variables satisfy $\bar e_j\ge p/(q^\star p_1)$. Consequently at least $p_1$ of the $p$ averaged e-values are at least $p/(q^\star p_1)$, so the $p_1$-th largest order statistic obeys $\bar e_{(p_1)}\ge p/(q^\star p_1)$. The index $k=p_1$ therefore belongs to the set $\{k:\bar e_{(k)}\ge p/(q^\star k)\}$, and by maximality in \eqref{eq:ebh},
    \begin{equation*}
        \widehat{k}\ge p_1.
    \end{equation*}
    Hence $p/(q^\star\widehat{k})\le p/(q^\star p_1)\le\bar e_j$ for every $j\in \mathcal{H}_1$, that is $\mathcal{H}_1\subseteq\widehat{\mathcal{S}}$. Since $\mathbb{P}_\mathbf{w}(\mathcal{E}\cap\mathcal{D})\to1$ by \eqref{eq:E-union}, this proves the sure-screening recovery.

    Finally, the ratio $|\widehat{\mathcal{S}}\cap \mathcal{H}_1|/p_1$ lies in $[0,1]$ always and equals one on $\{\mathcal{H}_1\subseteq\widehat{\mathcal{S}}\}$, so $1\ge\Pi^{\text{e-BH}}\ge\mathbb{P}_\mathbf{w}\bigl(\mathcal{H}_1\subseteq\widehat{\mathcal{S}}\bigr)\to1$, giving $\Pi^{\text{e-BH}}\to1$.
\end{proof}

\begin{proof}[Proof of Proposition~\ref{prop:hd-power}]
    Condition on $\widehat{J}$ and work on $\mathcal{F}_p$. By (S1), $\mathcal{H}_1\subseteq\widehat{J}$, so $\hat p_1=p_1$ and the reduced model $y^{(2)}=X^{(2)}_{\widehat{J}}\beta^0_{\widehat{J}}+\epsilon^{(2)}$ is correctly specified with active set $\mathcal{H}_1$. As established in the proof of Proposition~\ref{prop:hd}, the geometry of Section~\ref{sec:bm_variables}, the symmetry of Lemma~\ref{lemma1}, and conditions (C1)--(C4) all hold for the screened problem under (S2)--(S3). The derivations of Section~\ref{sec:per-iterate-power} use only these, together with (A1), and the decomposition \eqref{eq:d-comp} of $d_j^\pm$ under the alternative, which is a statement about the reduced model. Hence Lemma~\ref{lem:power-std}, Theorem~\ref{th:5} and Corollary~\ref{cor:power-global} hold with $\hat r,\hat p_1, \hat\psi_j$ in place of $r,p_1,\psi_j$, and Theorem~\ref{th:6} follows under \eqref{eq:alpha-condition} and \eqref{eq:T-condition}, giving \eqref{eq:hd-sure-discovery} conditionally on $\widehat{J}$. Variables outside $\widehat{J}$ have $\widetilde M_j^{(t)}=0<\tau_q^{(t)}$ and are never selected, but by (S1) none of them is active, so $\widehat{\mathcal{S}}\cap\mathcal{H}_1$ is unaffected. Finally, the ratio in \eqref{eq:hd-sure-discovery} lies in $[0,1]$ and the conditional bound holds with the same constants for every realisation of $\widehat{J}$ in $\mathcal{F}_p$, so the tower property gives the unconditional statements at an additive cost of $\mathbb{P}(\mathcal{F}_p^c)\to0$.
\end{proof}

\bibliographystyle{apalike}
\bibliography{refs}

\end{document}